\documentclass[aps,pra,preprint,superscriptaddress]{revtex4}
\usepackage{amsmath,epsfig}
\usepackage{graphicx}

\def\be{\begin{equation}}
\def\ee{\end{equation}}
\def\bea{\begin{eqnarray}}
\def\eea{\end{eqnarray}}
\newcommand{\ket}[1]{\mbox{$|#1\rangle$}}

\newcommand{\avg}[1]{\mbox{$\langle#1\rangle$}}

\def\bfr{{\bf r}}
\def\bfrp{{\bf {r^\prime}}}

\def\bfB{{\bf B}}
\def\bfE{{\bf E}}

\def\bfk{{\bf k}}

\newcommand{\opdagger}[2]{\mbox{$\hat{#1}_{#2}^{\dagger}$}}
\newcommand{\op}[2]{\mbox{$\hat{#1}_{#2}$}}
\def\fancyF{\mathcal{F}}

\begin{document}
\title{Cavity optomechanics using an optically levitated nanosphere}

\author{D.E. Chang}
\affiliation{Institute for Quantum Information and Center for the Physics of Information, California Institute of Technology, Pasadena, CA 91125}

\author{C.A. Regal}
\affiliation{Norman Bridge Laboratory of Physics 12-33, California Institute of Technology, Pasadena, CA 91125}

\author{S.B. Papp}
\affiliation{Norman Bridge Laboratory of Physics 12-33, California Institute of Technology, Pasadena, CA 91125}

\author{D.J. Wilson}
\affiliation{Norman Bridge Laboratory of Physics 12-33, California Institute of Technology, Pasadena, CA 91125}

\author{J. Ye}
\affiliation{Norman Bridge Laboratory of Physics 12-33, California
Institute of Technology, Pasadena, CA 91125} \affiliation{JILA,
NIST, and Department of Physics, University of Colorado, Boulder,
CO 80309}

\author{O.J. Painter}
\affiliation{Department of Applied Physics, California Institute of Technology, Pasadena, CA 91125}

\author{H.J. Kimble}
\affiliation{Norman Bridge Laboratory of Physics 12-33, California Institute of Technology, Pasadena, CA 91125}

\author{P. Zoller}
\affiliation{Norman Bridge Laboratory of Physics 12-33, California
Institute of Technology, Pasadena, CA 91125}
\affiliation{Institute for Quantum Optics and Quantum Information
of the Austrian Academy of Sciences, A-6020 Innsbruck, Austria}

\date{\today}

\begin{abstract}

Recently, remarkable advances have been made in coupling a number
of high-Q modes of nano-mechanical systems to high-finesse optical
cavities, with the goal of reaching regimes where quantum behavior
can be observed and leveraged toward new applications. To reach
this regime, the coupling between these systems and their thermal
environments must be minimized. Here we propose a novel approach
to this problem, in which optically levitating a nano-mechanical
system can greatly reduce its thermal contact, while
simultaneously eliminating dissipation arising from clamping.
Through the long coherence times allowed, this approach
potentially opens the door to ground-state cooling and coherent
manipulation of a single mesoscopic mechanical system or
entanglement generation between spatially separate systems, even
in room temperature environments. As an example, we show that
these goals should be achievable when the mechanical mode consists
of the center-of-mass motion of a levitated nanosphere.

\end{abstract}

\maketitle

One of the most intriguing questions associated with quantum
theory is whether effects such as quantum coherence and
entanglement can be observed at mesoscopic or macroscopic scales.
As a first step towards resolving this question, recently much
effort has been directed toward quantum state preparation of a
number of high-Q modes of nano- and micro-mechanical systems -- in
particular, cooling such systems to their quantum ground
state~\cite{cleland09}. Reaching a regime where the quantum
behavior of mechanical devices emerges is not only of fundamental
interest, but could lead to new applications in fields such as
ultra-sensitive detection~\cite{mamin01} and quantum information
science~\cite{cleland04}. To reach this regime, it is critical
that the thermalization and decoherence rates of these systems be
minimized, by reducing the coupling to their thermal reservoirs.
Thus far, this has necessitated the use of cryogenic operating
environments. From an engineering standpoint, it would also be
desirable to reduce the dissipation and thermalization rates of
these systems through their clamping and material
supports~\cite{hao03}, so that these rates might approach their
fundamental material limits~\cite{lifshitz00}.

Here we propose a novel approach toward this problem, where the
material supports are completely eliminated by optically
levitating~\cite{ashkin07} a nano-mechanical system inside a
Fabry-Perot optical cavity. Indeed, since the pioneering work of
Ashkin on optical trapping of dielectric
particles~\cite{ashkin07}~(in the classical domain), it has been
realized that levitation under good vacuum conditions can lead to
extremely low mechanical damping
rates~\cite{ashkin76,libbrecht04}. We show, however, that such an
approach should also facilitate the emergence of quantum behavior
even in room-temperature environments, when the particles are of
sub-wavelength scale such that the effects of optical scattering
become negligible. As a specific example, we show that the
center-of-mass~(CM) motion of a levitated nanosphere can be
optically self-cooled~\cite{braginsky02,wilson-rae07,marquardt07}
to the ground state starting from room temperature. This system
constitutes an extreme example of environmental isolation, as the
CM motion is naturally decoupled from the internal degrees of
freedom in addition to the external isolation provided by
levitation. The long coherence time also allows for the
preparation of more exotic states through \textit{coherent}
quantum evolution. Here, we consider in detail two examples.
First, we describe a technique to prepare a squeezed motional
state, which can subsequently be mapped onto light leaving the
cavity using quantum state transfer
protocols~\cite{zeng94,parkins99,zhang03,jahne09}. Under realistic
conditions, the output light exhibits up to ${\sim}15$~dB of
squeezing relative to vacuum noise levels, potentially making this
system a viable alternative to traditional techniques using
nonlinear crystals~\cite{wu87}. Second, we show that entanglement
originally shared between two modes of light~\cite{yonezawa07} can
be efficiently transferred onto the motion of two spheres trapped
in spatially separate cavities, creating well-known
Einstein-Podolsky-Rosen~(EPR) correlations~\cite{einstein35}
between the mechanical systems. Our approach of optical levitation
mirrors many successful efforts to cool~\cite{mckeever03,maunz04},
manipulate~\cite{leibfried03} and entangle~\cite{jost09} the
motion of atoms and ions in room-temperature environments. At the
same time, our system has a number of potential advantages, in
that it enables direct imaging via strong fluorescence, exhibits
large trap depths, and has a relatively large mass. We also note
recent related experiments involving opto-mechanical
``fluids''~(with a continuous excitation spectrum rather than
discrete modes) in the form of trapped, ultracold atomic
gases~\cite{murch08,brennecke08}.

Beyond the examples presented here, the use of a levitated device
as an opto-mechanical system could open the door to many
interesting opportunities. For instance, it should allow
mechanical damping to approach fundamental material limits,
potentially enabling the exploration of nanoscale material
properties. By levitating systems with internal vibrational modes,
multiple modes could be optically addressed and cooled. In
addition, the CM oscillation frequency can be tuned through the
trapping intensity, allowing for adiabatic state
transfer~\cite{zener32} with other modes or matching spatially
separate systems for optical linking and entanglement
generation~\cite{cirac97}. Furthermore, this paradigm integrates
nano-mechanics with many techniques for atomic trapping and
manipulation, which can be further extended by levitating systems
containing an internal electronic transition~(\textit{e.g.}, a
color center within a nano-crystal~\cite{beveratos01}). Finally,
as illustrated by squeezed light generation, engineering
mechanical nonlinearities in conjunction with quantum state
transfer yields a novel means to realize nonlinear optical
processes.

\section{Optical forces and noise acting on a dielectric sphere}

\begin{figure}[p]
\begin{center}
\includegraphics[width=18cm]{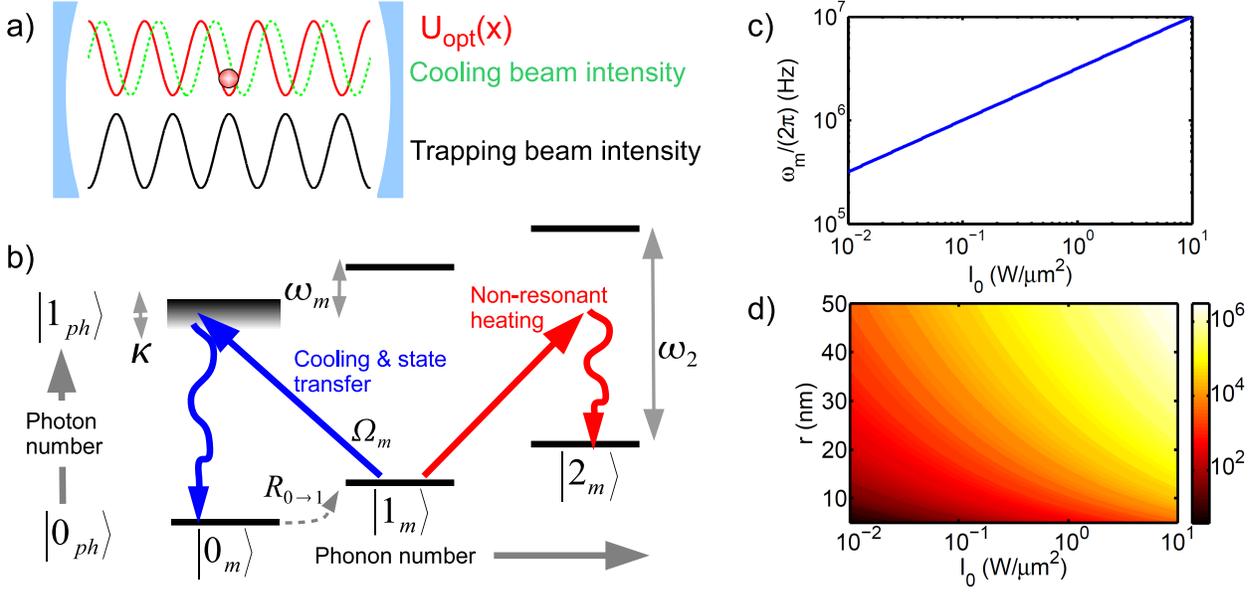}
\end{center}
\caption{a) Illustration of dielectric sphere trapped in optical
cavity. The large trapping beam intensity provides an optical
potential $U_{\footnotesize\textrm{opt}}(x)$ that traps the sphere
near an anti-node. A second more weakly driven cavity mode with a
non-vanishing intensity gradient at the trap center is used to
cool the motion of the sphere. b) Energy level diagram of
mechanical motion~(denoted $m$) and cavity cooling mode~(ph). The
mechanical mode has frequency $\omega_{m}$, while the optical mode
has frequency $\omega_2$ and linewidth $\kappa$. Photon recoil
induces transitions between mechanical states
$\ket{n_m}{\rightarrow}\ket{(n{\pm}1)_m}$ at a rate
$R_{n{\rightarrow}n{\pm}1}$~($R_{0{\rightarrow}1}$ shown by dashed
gray arrow). The cooling beam, with effective opto-mechanical
driving amplitude $\Omega_m$, induces anti-Stokes scattering that
cools the mechanical motion and allows for quantum state transfer
between motion and light. This beam is also responsible for
weaker, off-resonant heating via Stokes scattering. c) Mechanical
frequency $\omega_{m}$ as a function of trapping beam intensity.
For all numerical results, we take $\lambda=1\;\mu$m,
$\rho=2\;$g/cm${}^3$, and a high index
material~($\frac{\epsilon-1}{\epsilon+2}{\sim}1$). d) Optical trap
depth $U_0$~(in K) as functions of trapping beam intensity and
sphere radius $r$. \label{fig:schematic}}
\end{figure}

To illustrate our idea, we consider a sub-wavelength dielectric sphere interacting with two standing-wave optical modes of a Fabry-Perot cavity~(Fig.~\ref{fig:schematic}a). One resonantly driven mode provides an optical dipole trap for the sphere. The second mode is driven by a weaker ``cooling'' beam, assumed to have a non-zero intensity gradient at the trap center, which provides a radiation pressure cooling force~\cite{braginsky02,wilson-rae07,marquardt07}. We discuss the cooling mechanism in the next section, while here we focus on the trapping potential and the noise forces acting on the sphere.

The trapping beam provides a gradient force similar to that used
to ``optically tweeze'' small dielectric
particles~\cite{ashkin07}. Considering a sphere whose radius is
much smaller than the optical wavelength, $r{\ll}\lambda$, its
optical response is like that of a point dipole with induced
dipole moment
$p_{\footnotesize\textrm{ind}}={\alpha}_{\footnotesize\textrm{ind}}E(x)$
and optical potential
$U_{\footnotesize\textrm{opt}}(x)=-(1/4)(\textrm{Re}\;\alpha_{\footnotesize\textrm{ind}})|E(x)|^2$~(see
Appendix). Here $x$ is the CM position of the sphere,
$\alpha_{\footnotesize\textrm{ind}}=3\epsilon_{0}V\left(\frac{\epsilon-1}{\epsilon+2}\right)$
is its polarizability, $V$ is the sphere volume, and $\epsilon$ is
the electric permittivity. Taking a standing wave
$E(x)=E_{0}\cos\,kx$~($k{\equiv}2\pi/\lambda$), to lowest order
near an anti-node the potential corresponds to a harmonic
oscillator with mechanical frequency
\be \omega_{m}=\left(\frac{6k^{2}I_0}{\rho c}
\textrm{Re}\frac{\epsilon-1}{\epsilon+2}\right)^{1/2},\label{eq:omegam}
\ee
where $I_0$ is the field intensity and $\rho$ is the mass density
of the sphere. The total trap depth is
$U_{0}=(3I_{0}V/c)\textrm{Re}\frac{\epsilon-1}{\epsilon+2}$.
Typical trap depths and oscillation frequencies for a high-index
material~($\frac{\epsilon-1}{\epsilon+2}{\sim}1$) are plotted in
Figs.~\ref{fig:schematic}c,d. Frequencies of several MHz are
achievable using an intra-cavity intensity of
$I_{0}{\sim}1$~W/$\mu$m${}^2$. The imaginary component of
$\epsilon$ characterizes optical absorption, which contributes to
internal heating. For a material with ${\sim}10$~dB/km propagation
losses in bulk, intensities of $I_0{\sim}10$~W/$\mu$m${}^2$ can be
sustained without melting the sphere, due to blackbody
re-radiation of the absorbed energy~(see Appendix). With this in
mind, we assume $\epsilon$ is real in following discussions.

The dominant noise forces acting on the sphere are collisions with
a background gas and momentum recoil kicks due to scattered
photons. In the Appendix, we show that the contributions from shot
noise, blackbody radiation, and sphere anisotropy are negligible.
Furthermore, the CM is de-coupled from the internal degrees of
freedom and the sphere effectively has no internal structure~(as
opposed to molecules, where the internal configuration can affect
cooling efficiency~\cite{bahns96}). In the regime where the
molecular mean free path exceeds $r$, the background gas leads to
a mean damping force $dp/dt=-\gamma_{g}p/2$ with damping rate
$\gamma_{g}/2=(8/\pi)(P/\bar{v}r\rho)$, where $P,\bar{v}$ are the
background gas pressure and mean speed,
respectively~\cite{epstein24}. The random nature of the collisions
also thermalizes the motional energy, at a rate given through the
fluctuation-dissipation theorem by $dE/dt=-\gamma_{g}(E-k_{B}T)$,
where $T$ is the gas temperature. In particular, the
characteristic time for the system to heat by one phonon starting
from the ground state is
$\tau_{g}{\sim}\hbar\omega_{m}/\gamma_{g}k_{B}T$. Note that
$\tau_{g}^{-1}$ does not necessarily reflect the actual collision
rate between the sphere and gas molecules,
$R_{\footnotesize\textrm{coll}}{\approx}{\pi}P\bar{v}r^2/k_{B}T$~(it
is possible for a single collision to be quite rare,
$R_{\footnotesize\textrm{coll}}{\gg}\tau_{g}^{-1}$, and to impart
several phonons at once). We define a mechanical quality factor
$Q_g=\omega_m/\gamma_g$ due to the background gas, and a number of
coherent oscillations
$N^{(g)}_{\footnotesize\textrm{osc}}\equiv\omega_{m}\tau_{g}/2\pi$
expected before the energy increases by a single phonon. For a
sphere of radius $r=50$~nm, $\omega_{m}/(2\pi)=1$~MHz, and a
room-temperature gas with $P=10^{-10}$~Torr, one finds
$\gamma_{g}{\sim}10^{-6}$~s${}^{-1}$,$Q_g{\sim}6{\times}10^{12},N^{(g)}_{\footnotesize\textrm{osc}}{\sim}10^5$,
indicating that the levitated sphere can be essentially
\textit{de-coupled} from its thermal environment.

Photons scattered by the sphere out of the cavity lead to heating
via momentum recoil kicks. In analogy with atoms or ions trapped
in the Lamb-Dicke regime~\cite{leibfried03}~(when the particle is
trapped on a scale $\Delta{x}$ much smaller than $\lambda$), the
scattering induces transitions between consecutive harmonic
oscillator levels $n{\rightarrow}n{\pm}1$, with rates
$R_{n{\rightarrow}n{\pm}1}=\gamma_{\footnotesize\textrm{sc}}(n+1/2{\pm}1/2)$.
Second-order perturbation theory~\cite{wineland79} yields
\be
\gamma_{\footnotesize\textrm{sc}}=(2/5)(\omega_{r}/\omega_{m})R_{\footnotesize\textrm{sc}},\label{eq:gammasc}
\ee
where $\omega_{r}=\hbar k^2/2{\rho}V$ is the recoil frequency and $R_{\footnotesize\textrm{sc}}=48\pi^{3}\frac{I_{0}V^2}{\lambda^{4}\hbar\omega}(\frac{\epsilon-1}{ \epsilon+2})^2$ is the photon scattering rate. A result identical to Eq.~(\ref{eq:gammasc}) holds for a weakly excited, trapped atom~\cite{cirac92}. When photon scattering dominates the heating, the expected number of coherent oscillations is
\be
N^{(\footnotesize\textrm{sc})}_{\footnotesize\textrm{osc}}\equiv\frac{\omega_m}{2\pi\gamma_{\footnotesize\textrm{sc}}}=\frac{5}{
8\pi^3}\frac{\epsilon+2}{\epsilon-1}\frac{\lambda^3}{V}.\label{eq:Nosc}
\ee
Note that $N^{(\footnotesize\textrm{sc})}_{\footnotesize\textrm{osc}}$ scales inversely with
the sphere volume~($N^{(\footnotesize\textrm{sc})}_{\footnotesize\textrm{osc}}{\sim}40$ for $r=50$~nm, $\lambda=1\;\mu$m, $\epsilon{\gg}1$), due to the fact that the scattered power and dipole force scale like $p_{\footnotesize\textrm{ind}}^2$ and $p_{\footnotesize\textrm{ind}}$, respectively. Comparing with background gas collisions at $P=10^{-10}$~Torr and $\omega_m/(2\pi)=1$~MHz, recoil heating dominates $N_{\footnotesize\textrm{osc}}$ for sphere sizes $r{\gtrsim}5$~nm. Reaching the regime $N_{\footnotesize\textrm{osc}}{\gg}1$ implies that the sphere can coherently evolve for many oscillation periods after any cooling mechanisms are turned off, which makes this system a promising candidate for observing coherent quantum phenomena.

Finally, we remark that $R_{\footnotesize\textrm{sc}}$ can be very
large~($R_{\footnotesize\textrm{sc}}{\sim}10^{15}$~s${}^{-1}$ for
$I_{0}=1$~W/$\mu$m${}^2$ and $r=50$~nm) compared to atoms or ions,
which enables direct imaging. The large scattering is due to the
large intensities and the linear response of the sphere~(it is not
saturated like an atom or ion), as opposed to the system behaving
as a lossy element in the cavity. The contribution to the cavity
loss rate is
$\kappa_{sc}=12\pi^{2}\omega(V^{2}/\lambda^{3}V_c)\left(\frac{\epsilon-1}{\epsilon+2}\right)^{2}$,
where $V_c$ is the cavity mode volume, and is typically much
smaller than the natural cavity linewidth $\kappa$. We also
emphasize that in the Lamb-Dicke regime, the scattering does not
cause extra decoherence beyond that from recoil heating. This is
in contrast to motional wavepackets of spatial extent
$\Delta{x}\sim\lambda$, where a single scattering event can
destroy quantum coherence~\cite{kokorowski01}.

\section{Cooling the center-of-mass motion to the ground state}

We now describe the optical cooling effect of the weaker, second cavity mode~(denoted mode $2$). For concreteness, we assume that the sphere is trapped near the anti-node $x=0$ of cavity mode $E_{1}{\propto}\cos\;k_{1}x$, and that the second mode has spatial profile $E_{2}{\propto}\cos\;(k_{2}x-\pi/4)$, such that the intensity gradient is maximized. The total Hamiltonian of the system is given in a
rotating frame by
\bea H & = &
-\hbar\delta_{1}\opdagger{a}{1}\op{a}{1}-\hbar\delta_{2}\opdagger{a}{2}\op{a}{2}
+\frac{\hbar\Omega}{2}\left[(\op{a}{1}+\opdagger{a}{1})+\sqrt{2\zeta'}(\op{a}{2}
+\opdagger{a}{2})\right]
\nonumber \\ & & -\hbar
g_{1}(\cos\;2k_{1}\hat{x}-1)\opdagger{a}{1}\op{a}{1}-\hbar
g_{2}\cos\;2(k_{2}\hat{x}-\pi/4)\opdagger{a}{2}\op{a}{2}+\frac{\hat{p}^2}{2m}
.\label{eq:H}
\eea
Here $\hat{p},\hat{x}$ are the momentum and position operators of
the CM, $\op{a}{i}$ is the photon annihilation operator of cavity
mode $i$, and $\Omega$, $\Omega\sqrt{2\zeta'}$ are the driving
amplitudes of modes $1$ and $2$, respectively. $\delta_{i}$ is the
detuning between the driving field and mode frequency when the
sphere sits at $x=0$. The opto-mechanical coupling strengths
$g_i=\frac{3V}{4V_{c,i}}\frac{\epsilon-1}{\epsilon+2}\omega_i$
characterize the position-dependent frequency shifts due to the
sphere~(see Appendix), where $V_{c,i}$, $\omega_i$ are the mode
volume and resonance frequency of mode $i$. To simplify notation,
we assume that modes $1,2$ have similar properties,
$\omega_{1}{\approx}\omega_{2}=\omega$, etc. In addition to the
evolution described by $H$, the system also exhibits cavity losses
and the mechanical noise described previously.

Expanding the opto-mechanical coupling term of mode $2$ around
$x=0$, $\hbar g
\cos\;2(k\hat{x}-\pi/4)\opdagger{a}{2}\op{a}{2}\approx 2\hbar g
k\hat{x} \opdagger{a}{2}\op{a}{2}$, one finds a linear coupling in
the sphere position, analogous to the effect of radiation pressure
on a moving mirror of a Fabry-Perot cavity~\cite{wilson-rae07}.
Physically, the motion induces changes in the detuning and thus
the intra-cavity field amplitude, while the lag in the cavity
response enables the field to do work~(cooling) on the sphere. To
calculate the cooling rate, following the techniques of
Ref.~\cite{wilson-rae07} we first apply shifts to the operators,
$\op{a}{i}\rightarrow\op{a}{i}+\alpha_i$,
$\hat{x}\rightarrow\hat{x}+x_0$, where $\alpha_i$ and
$x_0\approx\zeta/k$~($\zeta\approx\kappa^{2}\zeta'/(\kappa^{2}+4\delta_{2}^2)$)
are mean values of the quantum fields. Here we have defined
$2\zeta=|\alpha_{2}/\alpha_{1}|^2$ as the ratio of intra-cavity
intensities of modes $1$ and $2$, and assumed that mode $1$ is
driven on resonance~($\delta_1=0$). To lowest order in $\zeta$,
field mode $1$($2$) is purely responsible for trapping~(cooling).
Subsequently tracing out the cavity degrees of freedom yields
equations for the mechanical motion alone. In particular, to
lowest order in $\zeta$ and for $\delta_2<0$, the cooling laser
provides a net cooling rate $\Gamma\equiv
R_{opt,-}-R_{opt,+}=\kappa\Omega_{m}^2\left[((\delta_{2}+\omega_m)^2+(\kappa/2)^2)^{-1}-((\delta_{2}-\omega_m)^2+(\kappa/2)^2)^{-1}\right]$~(see
Appendix), where $R_{opt,{\mp}}$ denote the anti-Stokes~(cooling)
and Stokes~(heating) scattering rates~(see
Fig.~\ref{fig:schematic}b). Here $\Omega_{m}\equiv
2gkx_{m}|\alpha_{1}|\sqrt{2\zeta}$ is the effective
opto-mechanical driving amplitude~(see Fig.~\ref{fig:schematic}b)
and $x_m\equiv\sqrt{\hbar/2m\omega_m}$. Validity of these
perturbative results requires $\Omega_{m}\lesssim\kappa,\omega_m$
and $\zeta{\lesssim}1$.

In the realistic limit that background gas collisions are negligible, the steady-state phonon number is $\avg{n_f}{\approx}\tilde{n}_{f}+\gamma_{sc}/\Gamma$, where $\tilde{n}_{f}=R_{opt,+}/\Gamma$ is the fundamental limit of laser cooling~\cite{wilson-rae07}. It is minimized when $\delta_{2}=-(1/2)\sqrt{\kappa^2+4\omega_m^2}$. In particular, when sideband resolution is achieved~($\omega_m\gtrsim\kappa$), $\tilde{n}_{f,\footnotesize\textrm{min}}{\approx}(\kappa/4\omega_m)^2{\ll}1$, indicating that ground-state cooling is possible provided other heating mechanisms are made sufficiently small. Considering the limit $\omega_{m}{\gg}\kappa$ and taking the maximum cooling rate $\Gamma{\sim}\kappa$ consistent with the perturbative calculations, using Eq.~(\ref{eq:Nosc}) one can then re-write $\avg{n_f}$ as
\be
\avg{n_f}{\approx}\frac{\kappa^2}{16\omega_m^2}+\phi\frac{\omega_m}{\kappa}
.\;\;\;\;\;(\omega_{m}{\gg}\kappa)\label{eq:nf}
\ee
The last term on the right corresponds to photon recoil heating and $\phi=(4\pi^2/5)(V/\lambda^3)\frac{\epsilon-1}{\epsilon+2}$ is a dimensionless parameter characterizing the sphere volume. Eq.~(\ref{eq:nf}) is minimized for $\kappa/\omega_{m}=2\phi^{1/3}$, in which case $\avg{n_f}_{\footnotesize\textrm{min}}=3\phi^{2/3}/4{\propto}(r/\lambda)^{2}{\ll}1$. Thus, one sees that ground-state cooling is readily attainable~(provided that $\zeta{\lesssim}1$ can be simultaneously satisfied). Physically, the optimum value of $\kappa/\omega_m$ balances good sideband resolution and excessive heating when large intensities are used to increase $\omega_m$.

To illustrate these results, we consider a sphere of radius $r=50$~nm and $\omega_m/(2\pi)=0.5$~MHz levitated inside a cavity of length $L=1$~cm and mode waist $w=25\;\mu$m~($V_c=(\pi/4)Lw^2$). In Fig.~\ref{fig:cooling}a we plot the minimum obtainable $\avg{n_f}$~(black curve) as a function of cavity finesse $\fancyF\equiv{\pi}c/2{\kappa}L$, assuming negligible gas collisions and subject to the constraints $\zeta,\Omega_{m}/\kappa,\Omega_m/\omega_m<1/2$ and optimized over detuning $\delta_2$. For low cavity finesse the cooling is essentially limited by sideband resolution~($\tilde{n}_{f,\footnotesize\textrm{min}}$, red curve) and the ground state regime $\avg{n_f}{\sim}1$ can be reached with a finesse of $\fancyF{\sim}3600$. A minimum of $\avg{n_f}{\sim}0.02$ is reached at a finesse of $\fancyF{\sim}50000$~(with corresponding cooling rate $\Gamma{\sim}10^{6}$~s${}^{-1}$). This corresponds to a final temperature of $T_f{\sim}6\;\mu$K, or a remarkable compression factor of $T/T_f{\sim}5{\times}10^7$ relative to room temperature $T$.

\section{Motional entanglement and squeezed light generation using quantum state transfer}

A number of related schemes have been proposed for quantum state transfer between light and the motion of atoms~\cite{zeng94,parkins99} or nano-mechanical systems~\cite{zhang03,jahne09}. In our system, the small mechanical noise and ease of achieving good sideband resolution in principle allow state transfer to be realized with almost perfect efficiency. This might enable light with non-classical properties to be mapped onto mechanical motion, and as an example, we show that this can be used to generate EPR correlations between two spatially separate spheres. Moreover, a complementary process can be realized, where a non-trivial mechanical state~(a squeezed state) is prepared through coherent manipulation and subsequently transferred to light leaving the cavity. The latter case illustrates how opto-mechanics can yield a novel nonlinear optical system.

First we give a simplified picture of quantum state transfer using a one-sided, ideal cavity~(where all losses are via transmission through one cavity mirror)~\cite{gardiner85}. Specifically, we consider the Heisenberg equations of motion in a rotating frame for the cavity cooling mode and the motion~(after applying the shifts described in the previous section), when the cooling mode is driven resonantly on the red motional sideband~($\delta_{2}=-\omega_m$),
\bea \frac{d}{dt}\op{a}{2} & = &
-\frac{\kappa}{2}\op{a}{2}-i\Omega_{m}\left(\op{b}{}+\opdagger{b}{}e^{2i\omega_{
m}t}\right)+\sqrt{\kappa}\op{a}{2,\footnotesize\textrm{in}},
\nonumber \\ \frac{d}{dt}\op{b}{} & = &
(i/\hbar)[H_{e},\hat{b}]-i\Omega_{m}\left(\op{a}{2}+\opdagger{a}{2}e^{2i\omega_{m}t}
\right)+i\hat{F}(t)e^{i\omega_{m}t}.\label{eq:timeevolution}
\eea
The Hamiltonian $H_e$ describes any external forces or couplings applied to the sphere beyond those in Eq.~(\ref{eq:H}), $\hat{b}$ is the annihilation operator corresponding to a harmonic oscillator of mass $m$ and frequency $\omega_m$, and $\hat{a}_{2,\footnotesize\textrm{in}}$ is the cavity input operator associated with losses. $F(t)$ is the~(Hermitian) noise due to photon recoil, which has correlations $\avg{F(t)F(t')}=\phi\omega_{m}\delta(t-t')$, and we assume all other noise is negligible. Since the cavity trapping mode~($\hat{a}_1$) effectively provides a harmonic potential and can otherwise be ignored, for simplicity we will omit the subscript $2$ as we refer to the cooling mode in future discussions. Temporarily assuming that the non-secular terms~($e^{2i\omega_{m}t}$) can be ignored and that the mechanical motion evolves slowly on time scales compared to $1/\kappa$, one can adiabatically eliminate the cavity mode to yield $\hat{a}{\approx}-2i(\Omega_{m}/\kappa)\hat{b}+(2/\sqrt{\kappa})\op{a}{\footnotesize\textrm{in}}$, and $d\hat{b}/dt{\approx}(i/\hbar)[H_e,\hat{b}]-(\Gamma/2)\hat{b}-i\sqrt{\Gamma}\op{a}{in}+i\hat{F}(t)e^{i\omega_{m}t}$, where $\Gamma{\equiv}4\Omega_{m}^2/\kappa$ is the cavity-induced cooling rate in the weak-driving limit~($\Omega_{m}\lesssim\kappa$). The cavity output is related to the input and intra-cavity fields through $\op{a}{\footnotesize\textrm{out}}=\sqrt{\kappa}\hat{a}-\op{a}{\footnotesize\textrm{in}}$~\cite{gardiner85}, or $\op{a}{\footnotesize\textrm{out}}{\approx}-i\sqrt{\Gamma}\hat{b}+\op{a}{\footnotesize\textrm{in}}$, which states that the mechanical motion is mapped onto the outgoing light. Physically, the cooling process converts phononic excitations into photonic excitations that leak out of the cavity. Generally, two mechanisms will degrade state transfer. First, $\hat{F}$ adds extra noise to the ideal state that one is attempting to transfer, with a strength characterized by the small parameter $\phi$. Second, the non-secular terms contribute to Stokes scattering, destroying the perfect phonon-photon correspondence, with a strength that is expected to be proportional to $(\kappa/\omega_m)^2$. Given that $\phi,(\kappa/\omega_m)^2$ can be made small, nearly perfect state transfer is possible in principle. We illustrate this with two examples, entanglement transfer and squeezed light generation.

\subsection{Entanglement transfer}

Here we describe how EPR correlations shared between two modes of
light~\cite{yonezawa07} can be mapped to the motion of two spheres
trapped in spatially separate cavities. Specifically, we define
quadrature operators for the input light for each of the two
systems~(denoted $A,B$), given by
$X^{(j)}_{+,\footnotesize\textrm{in}}=(\hat{a}^{(j)}_{\footnotesize\textrm{in}}+\hat{a}^{(j)\dagger}_{\footnotesize\textrm{in}})$,
$X^{(j)}_{-,\footnotesize\textrm{in}}=(\hat{a}^{(j)}_{\footnotesize\textrm{in}}-\hat{a}^{(j)\dagger}_{\footnotesize\textrm{in}})/i$
for $j=A,B$. A similar set of operators
$X^{(j)}_{\pm,m},X^{(j)}_{\pm,\footnotesize\textrm{out}}$ can be
defined for the motion and output light, by replacing
$\hat{a}^{(j)}_{\footnotesize\textrm{in}}{\rightarrow}\hat{b}^{(j)},\hat{a}^{(j)}_{\footnotesize\textrm{out}}$,
respectively. Of particular interest is the case where the two
input fields exhibit broadband EPR correlations between them,
\be \avg{(X^{(A)}_{+,\footnotesize\textrm{in}}(\omega)+X^{(B)}_{+,\footnotesize\textrm{in}}(\omega))^2}/2=\avg{(X^{(A)}_{-,\footnotesize\textrm{in}}(\omega)-X^{(B)}_{-,\footnotesize\textrm{in}}(\omega))^2}/2=e^{-2R}<1.\label{eq:EPRstate} \ee
When the variances satisfy $e^{-2R}<1$, the two modes exhibit correlations below vacuum level and are entangled~\cite{duan00}~(for concreteness, we assume the other combinations of quadratures satisfy $\avg{(X^{(A)}_{\pm,\footnotesize\textrm{in}}(\omega){\mp}X^{(B)}_{\pm,\footnotesize\textrm{in}}(\omega))^2}/2=e^{2R}$). Such EPR correlations have been observed with light and in the internal degrees of freedom of atomic ensembles~\cite{julsgaard01}, but have yet to be demonstrated using mechanical systems.

\begin{figure}[p]
\begin{center}
\includegraphics[width=18cm]{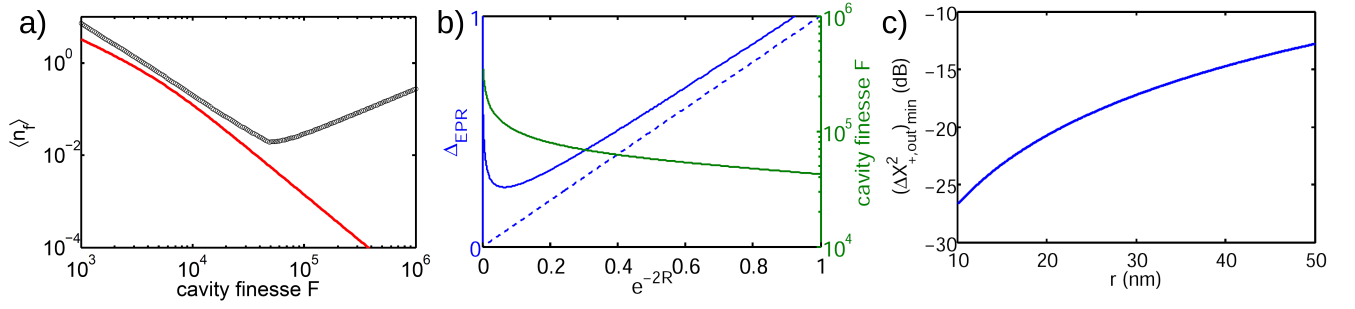}
\end{center}
\caption{a) Mean phonon number $\avg{n_f}$~(black curve) versus
cavity finesse $F$~($F={\pi}c/2{\kappa}L$) under optimized cooling
conditions. The sphere size is $r=50$~nm and the cavity has a
length and waist of $L=1$~cm, $w=25\;\mu$m, respectively. The red
curve denotes $\tilde{n}_{f,\footnotesize\textrm{min}}$, the
fundamental limit of cooling imposed by sideband resolution. b)
Solid blue curve: optimized EPR variance between two levitated
spheres, as a function of squeezing parameter $e^{-2R}$. System
parameters are identical to a). Dashed curve: EPR variance in
limit of perfect state transfer,
$\Delta_{\footnotesize\textrm{EPR}}=e^{-2R}$. Green curve: cavity
finesse corresponding to optimal EPR variance. c) Optimized
variance
$({\Delta}X_{+,\footnotesize\textrm{out}}^2)_{\footnotesize\textrm{min}}$~(in
dB) of squeezed output light from an ideal cavity, as a function
of sphere size.\label{fig:cooling}}
\end{figure}

To proceed, we solve Eq.~(\ref{eq:timeevolution}) in the Fourier
domain~(including the non-secular terms) for each of the systems
for the correlations given in Eq.~(\ref{eq:EPRstate}) and $H_e=0$.
Generally, the non-secular terms yield an infinite set of
algebraic equations~(coupling frequencies $\omega_m+2n\omega_{m}$
for integer $n$), which given $\omega_m{\gg}\Omega_m,\kappa$ can
be truncated to good approximation at $n>1$. For simplicity of
analysis, we assume the two systems have identical properties, and
that the cooling rate $\Gamma\sim\kappa$. However, we expect our
results should qualitatively hold provided that only
$\Gamma,\omega_m$ of the two systems are properly tuned, which can
be easily accomplished by adjusting the trapping and cooling beam
intensities. One can then show that state transfer yields the
following joint variances in the motion~(see Appendix),
\be \Delta_{\footnotesize\textrm{EPR}}\equiv\avg{(X^{(A)}_{\pm,m}(t){\mp}X^{(B)}_{\pm,m}(t))^2}/2=e^{-2R}+\frac{\kappa^2}{16\omega_m^2}(3e^{2R}+2\,{\sinh}\,2R)+\frac{4\phi\omega_m}{\kappa}. \ee
As expected, Stokes scattering and recoil heating contribute to
the variance by amounts $(\kappa/\omega_m)^2$ and
$\phi\omega_m/\kappa$, respectively. This can be minimized with
respect to $\kappa/\omega_m$, yielding
$\Delta_{\footnotesize\textrm{EPR,min}}=e^{-2R}+3(\phi/2)^{2/3}(3e^{2R}+2\sinh\,2R)^{1/3}$.
To illustrate these results we plot
$\Delta_{\footnotesize\textrm{EPR,min}}$ in
Fig.~\ref{fig:cooling}b as a function of $e^{-2R}$, taking the
same parameters as in Fig.~\ref{fig:cooling}a. For the moderate
values of $e^{-2R}$ typically obtained in
experiments~\cite{yonezawa07}, EPR correlations in the motion can
be achieved with reasonable cavity finesse $F<10^5$.

\subsection{Squeezed light generation}

First we describe a technique to create a mechanical squeezed
state, and then derive the properties of the outgoing light upon
quantum state transfer. Mechanical squeezing is accomplished by
adding a sinusoidally-varying component to the intensity of the
trapping beam, which yields the Hamiltonian of a parametric
amplifier~\cite{rugar91},
$H_{e}=\epsilon_{m}\omega_m^{2}x^2\sin\;2\omega_{m}t$. Here
$\epsilon_m$ is a small parameter characterizing the strength of
the modulation of the trap frequency. As one approaches the
threshold for parametric
oscillation~($\epsilon_{m}\omega_{m}{\rightarrow}\Gamma$), the
variance in one quadrature of motion is reduced by up to a factor
of $2$~\cite{rugar91}.

We now investigate the properties of the outgoing light over a
narrow frequency range near the cavity resonance, specifically
considering $X_{{\pm},\footnotesize\textrm{out}}(\omega=0)$. We
apply similar methods as above to solve
Eq.~(\ref{eq:timeevolution}) in the Fourier domain. Taking the
limit as one approaches threshold and $\Gamma{\sim}\kappa$, the
variance in the output light is given by~(see Appendix)
\be
{\Delta}X_{+,\footnotesize\textrm{out}}^2(\omega=0)=\frac{2\phi\omega_m}{\kappa}+\frac{5}{16}\frac{\kappa^2}{\omega_m^2}.\label{eq:squeezing}
\ee
Again, an optimum value of $\kappa/\omega_m{\propto}\phi^{1/3}$
maximizes the squeezing, with
$({\Delta}X^{2}_{+,\footnotesize\textrm{out}})_{\footnotesize\textrm{min}}{\approx}2.04\phi^{2/3}$~(note
that ${\Delta}X^{2}_{+,\footnotesize\textrm{out}}=1$ for vacuum).
A plot of
$({\Delta}X^{2}_{+,\footnotesize\textrm{out}})_{\footnotesize\textrm{min}}$
as a function of sphere size is shown in Fig.~\ref{fig:cooling}c.
For $r=10$~nm size spheres, one finds that over $25$~dB of
squeezing relative to vacuum can be obtained using an ideal
cavity~(note for good vacuum conditions, background gas collisions
are negligible down to $r{\sim}5$~nm). In practice, a cavity has
additional scattering and absorption losses that limit the
squeezing. Taking an ultra-high finesse cavity with ${\sim}1$~ppm
losses per round trip~\cite{hood01} and a set of reasonable cavity
dimensions, we show in the Appendix that light squeezed by up to
${\sim}15$~dB can be extracted.

In principle, similar techniques also apply to trapped atoms or
ions. However, one benefits from the relatively large mass $m$ of
the sphere. Specifically, approaching threshold, one quadrature of
motion becomes infinitely unsqueezed, producing a large position
uncertainty ${\Delta}x$~\cite{rugar91}. At the same time, faithful
quantum state transfer requires a linear opto-mechanical coupling,
which translates into a requirement that the Lamb-Dicke parameter
$\eta{\equiv}k{\Delta}x\propto{m^{-1/2}}{\ll}1$ be small. In the
Appendix, we show that $\eta{\ll}10^{-2}$ can be satisfied with a
sphere even in the regime of ${\sim}25$~dB squeezing levels. To
compare, a typical atom trapped with a frequency of
$\omega_{m}/(2\pi){\sim}1$~MHz in its \textit{ground state}
already yields $\eta{\sim}0.05$.

\section{Outlook}

An optically levitated opto-mechanical system can have remarkably
long coherence times, which potentially enables quantum phenomena
such as entanglement to be observed even in room-temperature
environments. Combining previously demonstrated techniques to
controllably grow small particles~\cite{venkatathri08} and load
and manipulate them in vacuum~\cite{shu06,ashkin07} should put
this system within experimental reach. Extending the ideas
presented here should open up several other interesting
possibilities. For example, beyond the dipole-type objects
considered here, one could engineer the shapes of the levitated
objects to yield even larger mechanical frequencies and coherence
times, and controllably study the decoherence of a large
system~\cite{hackermuller04}. Also, several spheres or more
complex nano-mechanical systems with internal modes could be
levitated and coupled together, for the purpose of entangling
multiple degrees of freedom.  Separately, one could take advantage
of the fact that the potential for the CM is purely optical to
engineer non-trivial dynamics, such as nonlinear motion. It would
also be interesting to develop analogous levitation techniques
using nano- and micro-photonic
cavities~\cite{vahala03,vuckovic03}, combining their remarkable
optical properties with the excellent mechanical characteristics
of a trapped particle. Finally, by levitating charged or magnetic
systems, one could potentially realize systems analogous to ion
traps~\cite{wineland07} or facilitate novel quantum hybrid
architectures~\cite{rabl09}.

\acknowledgments

DC and SP acknowledge support from the Gordon and Betty Moore
Foundation through Caltech's Center for the Physics of
Information, DC from the National Science Foundation under Grant
No. PHY-0803371, CR from a Millikan Postdoctoral Fellowship, and
JY and PZ from a Moore Fellowship during their stay at Caltech.
Work at Innsbruck is supported by the Austrian Science Fund and EU
Projects.

Note added: We also have become aware of a recent, similar
proposal to optically levitate and manipulate a nano-mechanical
system by O. Romero-Isart \textit{et al}., in arXiv:0909.1469
(2009).

\appendix
\section{Electrodynamic calculation of forces on dielectric sphere}

Here we describe in detail the point-dipole approximation for a
sub-wavelength dielectric sphere, and compare the results of this
approximation with exact numerical electrodynamic calculations of
the optical forces. For concreteness, we consider a dielectric
sphere of (possibly complex) permittivity $\epsilon$ and radius
$r$, interacting with an incident standing electromagnetic wave
with electric and magnetic field components
$\bfE_{\footnotesize\textrm{in}}=\hat{x}E_{0}\cos\,k(z-z_0)\cos\,{\omega}t$
and
$\bfB_{\footnotesize\textrm{in}}=\hat{y}(E_{0}/c)\sin\,k(z-z_0)\sin\,{\omega}t$~($k=\omega/c$).
Here we assume that the sphere is in free space rather than in a
cavity, which allows one to unambiguously calculate the optical
forces acting on the sphere independent of its motion~(as opposed
to the cavity case where the motion of the sphere generally shifts
the cavity resonance, causing the intra-cavity field $E_{0}(t)$ to
depend on the history of motion). In the special case where the
sphere is localized near one of the nodes or anti-nodes, the
free-space and cavity cases yield the same results~(\textit{e.g.},
for the mechanical trap frequency $\omega_m$) as the cavity
resonance and intra-cavity field to lowest order become
insensitive to small displacements of the sphere. The
electrodynamic problem of plane-wave scattering off of a sphere is
exactly solvable, as the vector wave equation
$\nabla^{2}\bfE(\bfr)+k^{2}\epsilon(\bfr)\bfE(\bfr)=0$~(with
similar equation for $\bfB$) admits solutions through separation
of variables~\cite{stratton41}. Note that one can define natural
length scales $k|\sqrt{\epsilon}|r$, $kr$ for the electrodynamic
response inside and outside the sphere. Of particular interest is
the case when $k|\sqrt{\epsilon}|r{\ll}1$ is a small parameter~(we
assume that $|\sqrt{\epsilon}|>1$ for this discussion, which is
typically the case). One can then formally solve the wave
equations using perturbation theory, with the lowest order
equation given by $\nabla^{2}\bfE(\bfr)=0$ along with appropriate
boundary conditions at the surface of the sphere. Physically, this
approximation states that the magnetic field is not important in
the near-field, such that the lowest-order response of the sphere
can be obtained by solving an electrostatic equation. Taking an
optical wavelength of $\lambda=2\pi/k=1\;\mu$m and $\epsilon=2$,
for instance, the electrostatic solution should be valid for
$r{\lesssim}110$~nm. In this regime, the polarizablity of the
sphere is of the simple form given by electrostatic theory,
$\alpha_{\footnotesize\textrm{ind}}=3\epsilon_{0}V\frac{\epsilon-1}{\epsilon+2}$~\cite{stratton41},
as is used in the main text. The optical potential experienced by
the sphere is predicted to be
$U_{\footnotesize\textrm{opt}}=-(1/4)(\textrm{Re}\,\alpha_{\footnotesize\textrm{ind}})E_{0}^2\cos^{2}k(z-z_0)$.
For spheres larger than $r{\gtrsim}1/k|\sqrt{\epsilon}|$, the
forces predicted by the electrostatic theory will be substantially
larger than the actual forces, as phase variations of the field
within the sphere become important.

\begin{figure}[p]
\begin{center}
\includegraphics[width=10cm]{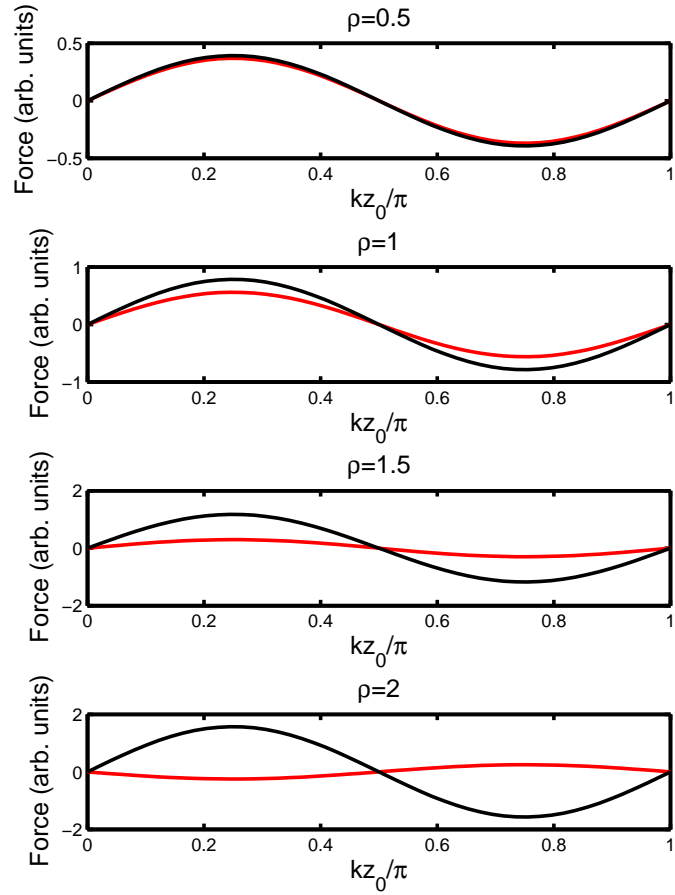}
\end{center}
\caption{Comparison of optical forces~(in arbitrary units) acting
on a dielectric sphere of permittivity $\epsilon=2$ as a function
of position $kz_0$. The four figures shown correspond to sphere
sizes of $\rho{\equiv}k\sqrt{\epsilon}r=0.5,1,1.5,2$. The black
curve indicates the results obtained from an electrostatic,
point-dipole approximation, while the red curve denotes exact
electrodynamical theory.\label{fig:forceplot}}
\end{figure}

To compare the electrostatic approximation with actual results, we
first solve the electrodynamic scattering problem exactly. The
exact force $F_z$ along $z$ can then be obtained by integrating
the Maxwell stress tensor $T_{ij}$ over the sphere surface $S$,
\be
F_{z}=\epsilon_{0}\oint_{S}\,da\,\sum_{j=x,y,z}T_{zj}\hat{n}_{j},
\ee
where $\hat{n}_j$ is the outgoing normal vector to the sphere
surface. In Fig.~\ref{fig:forceplot} we compare the approximate
and exact forces for various values of $r$, taking $\epsilon=2$.
It can be seen that the two methods agree closely for
$k\sqrt{\epsilon}r{\lesssim}1$. For spheres where
$k\sqrt{\epsilon}r{\gtrsim}1$, the forces predicted from
electrostatic theory can be much larger than the actual forces,
and even different in sign.

\section{Absorption losses of trapped sphere}

In this section, we consider the effect that a small imaginary
component of the permittivity $\epsilon$ has on a trapped sphere.
In the limit that the sphere has a radius much smaller than the
optical wavelength, the sphere behaves as a point-like dipole with
polarizability
$\alpha_{\footnotesize\textrm{ind}}=3\epsilon_{0}V\left(\frac{\epsilon-1}{\epsilon+2}\right)$.
For small Im~$\epsilon$, the polarizability acquires a small
imaginary component that leads to a non-zero absorption
cross-section, with a corresponding absorbed power
\be
P_{\footnotesize\textrm{abs}}=12\pi\frac{I_0}{\lambda}V\textrm{Im}\,\frac{\epsilon-1}{\epsilon+2}.
\ee
Here $I_0$ is the trapping beam intensity, $V$ is the sphere
volume, and $\lambda$ is the optical wavelength. The absorbed
power causes a rise in the internal temperature
$T_{\footnotesize\textrm{int}}$ of the sphere, which is balanced
out by thermalization with a background gas and blackbody
radiation.

We first quantify the effect of the background gas~(which is
negligible in the regime of particular interest where the sphere
is trapped under good vacuum conditions). There are two limiting
regimes to the background gas interactions, where the sphere
radius is either much smaller or larger than the molecular mean
free path $\lambda_{\footnotesize\textrm{mfp}}$. At a relatively
large pressure of $P=1$~Torr and room temperature,
$\lambda_{\footnotesize\textrm{mfp}}{\sim}100\;\mu$m and thus our
case of interest is always
$r{\ll}\lambda_{\footnotesize\textrm{mfp}}$. Here, gas molecules
independently collide and partially thermalize with the sphere.
This leads to a cooling rate~\cite{liu06}
\be
\frac{dE}{dt}=-\alpha_{g}\sqrt{\frac{2}{3\pi}}(\pi{r^2})Pv_{\footnotesize\textrm{rms}}\frac{\gamma_{\footnotesize\textrm{sh}}+1}{\gamma_{\footnotesize\textrm{sh}}-1}\left(\frac{T_{\footnotesize\textrm{int}}}{T}-1\right),
\ee
where $P,v_{\footnotesize\textrm{rms}},T$ are the background gas
pressure, root-mean-square speed, and temperature, respectively,
and $\gamma_{\footnotesize\textrm{sh}}$ is the gas specific heat
ratio~($\gamma_{\footnotesize\textrm{sh}}=7/5$ for an ideal
diatomic gas). $\alpha_g$ is a phenomenological energy
accommodation factor~($0\leq\alpha_g{\leq}1$), which characterizes
the degree to which a gas molecule thermalizes with the sphere
upon a single collision.

Under good vacuum conditions, blackbody radiation dissipates the
majority of the power absorbed by the sphere. For the sub-micron
spheres we are considering, the radius is much smaller than the
absorption length at typical blackbody radiation wavelengths, and
thus the usual formulas for blackbody radiated power do not apply.
Instead, the sphere again behaves as a point-like dipole at these
wavelengths, \textit{e.g.}, the radiated power scales like
volume~(as opposed to surface area in the case of a large object).
The internal heating rate due to blackbody radiation is given by
$dE/dt=\sum_{\bfk}({\hbar}ck)R_{\footnotesize\textrm{abs},\bfk}$,
where the sum is over all blackbody radiation modes~(and
polarizations), $\bfk$ is the wavevector of each mode, and
$R_{\footnotesize\textrm{abs},\bfk}$ is the absorption rate of
each mode. It is given by
\be
R_{\footnotesize\textrm{abs},\bfk}=3ck(V/V_q)n_{k}\textrm{Im}\left(\frac{\epsilon(\omega_k)-1}{\epsilon(\omega_k)+2}\right),\label{eq:Rabs}
\ee
where $n_k=(e^{{\hbar}ck/k_{B}T}-1)^{-1}$ is the occupation number
of each mode and $V_q$ is the quantization volume. Assuming that
the sphere has a relatively constant and temperature-independent
permittivity
$\epsilon(\omega){\approx}\epsilon_{\footnotesize\textrm{bb}}$
across the blackbody radiation spectrum, it is straightforward to
show that the sphere absorbs blackbody radiation at a rate
\be
\frac{dE}{dt}=\frac{72\zeta(5)}{\pi^2}\frac{V}{c^{3}\hbar^4}\textrm{Im}\,\left(\frac{\epsilon_{\footnotesize\textrm{bb}}-1}{\epsilon_{\footnotesize\textrm{bb}}+2}\right)(k_{B}T)^5,\label{eq:bbabs}
\ee
where $T$ is the background temperature and
$\zeta(5){\approx}1.04$ is the Riemann zeta function. Similarly,
the sphere radiates blackbody energy at a rate given by the
negative of Eq.~(\ref{eq:bbabs}), with the substitution
$T{\rightarrow}T_{\footnotesize\textrm{int}}$.

\begin{figure}[p]
\begin{center}
\includegraphics[width=18cm]{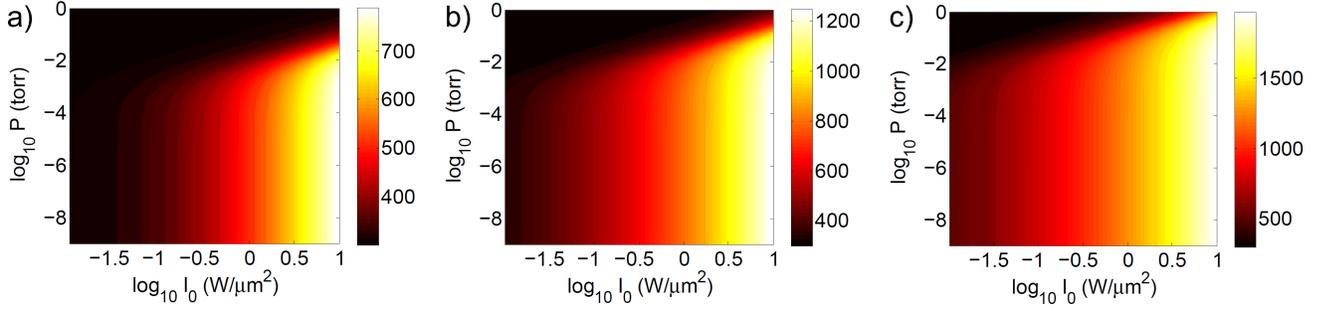}
\end{center}
\caption{Internal temperature of sphere~(in K), as functions of
background gas pressure and intra-cavity intensity. Material
parameters for the sphere are given in the text. Optical losses
for the sphere are assumed to be a) 10 dB/km, b) 100 dB/km, and c)
1000 dB/km.\label{fig:sphereheating}}
\end{figure}

To illustrate these results, in Figs.~\ref{fig:sphereheating}a-c
we plot the internal equilibrium temperature
$T_{\footnotesize\textrm{int}}$ of the sphere as a function of
background gas pressure and trapping intensity $I_0$. Here we have
taken into account the effects of optical
absorption~(Im~$\epsilon$), thermalization with the background
gas, and blackbody radiation. The values of Im~$\epsilon$ in
Figs.~\ref{fig:sphereheating}a,b,c correspond to bulk optical
absorption rates of $10,100,1000$~dB/km, respectively, while the
real part of the permittivity is chosen to be Re~$\epsilon=2$. We
have taken the other parameters to be $r=50$~nm, $\alpha_g=0.25$,
Im~$\frac{\epsilon_{\footnotesize\textrm{bb}}-1}{\epsilon_{\footnotesize\textrm{bb}}+2}=0.1$~(roughly
corresponding to the averaged value of fused silica around
blackbody wavelengths~\cite{lang83}), and a volumetric heat
capacity of the sphere of $\tilde{c}=2$~J/m${}^{3}\cdot$K. Note
that at sufficiently low pressures, the temperature becomes
pressure-independent as only blackbody radiation significantly
contributes to energy dissipation~(as indicated by the vertical
contours in the figure). Furthermore, in this regime the final
temperature is independent of the sphere size~(provided that
$r{\ll}\lambda$), since both the optical absorption and blackbody
radiation scale linearly with volume. For losses of
${\sim}10$~dB/km, one finds that over $10$~W/$\mu$m${}^2$ of power
can be sustained without exceeding the melting point of a typical
material.

\section{Derivation of opto-mechanical coupling strength}

Generally, introducing a dielectric material into an optical
cavity causes the bare resonant frequency $\omega$ of a cavity
mode to shift by an amount $\delta\omega$, which in perturbation
theory is given by~\cite{rodriguez07}
\be
\frac{\delta\omega}{\omega}=-\frac{1}{2}\frac{\int\;d^{3}\bfr\;\delta{P}(\bfr)\cdot\bfE(\bfr)}{\int\;d^{3}\bfr\;\epsilon_{0}\bfE^{2}(\bfr)}.\label{eq:cavityshift}
\ee
Here $\bfE(\bfr)$ is the bare cavity mode profile and
${\delta}P(\bfr)$ is the variation in permittivity introduced by
the dielectric object. Considering the case where the dielectric
object is a sub-wavelength sphere, its dielectric response is
well-approximated by a point dipole,
$P(\bfrp){\approx}\alpha_{\footnotesize\textrm{ind}}E(\bfr)\delta(\bfr-\bfrp)$,
where $\bfr$ is the center-of-mass~(CM) position of the sphere.
Taking a mode profile $E{\propto}\cos\;(kx-\phi)$, one readily
finds~(up to a constant shift) that
\be
\delta\omega=-\frac{3V}{4V_c}\frac{\epsilon-1}{\epsilon+2}\cos\;(2kx-2\phi)\omega.\label{eq:cmshift}
\ee
The interaction Hamiltonian between this optical mode and the
mechanical motion is subsequently given by
$H_{\footnotesize\textrm{om}}=\hbar\delta\omega\opdagger{a}{}\hat{a}$,
and as in the main text, one can define a characteristic
opto-mechanical coupling strength
$g=\frac{3V}{4V_c}\frac{\epsilon-1}{\epsilon+2}\omega$.

\section{Optical self-cooling equations}

Here we derive in detail the cooling rate equations for the CM
motion of the sphere, whose results are summarized in the main
text.  We begin with the Hamiltonian given by Eq.~(4) in the main
text. The corresponding Heisenberg equations of motion, including
dissipation, are
\bea \frac{d}{dt}\op{a}{1} & = &
(i\delta_{1}-\kappa/2)\op{a}{1}-\frac{i\Omega}{2}+\sqrt{\kappa}\op{a}{1,\footnotesize\textrm{in}},
\nonumber
\\ \frac{d}{dt}\op{a}{2} & = &
(i(\delta_{2}+2gk\hat{z})-\kappa/2)\op{a}{2}-\frac{i\Omega}{2}\sqrt{2\zeta'}+\sqrt{\kappa}\op{a}{2,\footnotesize\textrm{in}},
\nonumber
\\ \frac{d}{dt}\hat{p} & = &
-4{\hbar}gk^{2}\opdagger{a}{1}\op{a}{1}\hat{z}+2{\hbar}gk\opdagger{a}{2}\op{a}{2}-\gamma\hat{p}/2+\hat{F}_{p}(t),
\nonumber
\\ \frac{d}{dt}\hat{x} & = & \frac{\hat{p}}{m}.\label{eq:heisenbergeqs} \eea
Here $\op{a}{i,\footnotesize\textrm{in}}$ are input-field
operators associated with the cavity mode losses $\kappa$,
$\gamma$ is the damping rate of the motion, and $\hat{F}_p$ is the
noise force acting on the sphere. In the above equations, we have
expanded the position-dependent opto-mechanical coupling terms
$g_{i}\cos\,2(k_{i}\hat{x}-\phi_{i})$ to first order in the
displacement $\hat{x}$, and for simplicity have assumed that the
two cavity modes have similar properties~($g_1{\approx}g_2=g$,
etc.). We now apply shifts to all of the operators,
$\op{a}{i}{\rightarrow}\op{a}{i}+\alpha_{i}$,
$\hat{x}{\rightarrow}\hat{x}+x_0$, where the constants $x_{0}$ and
$\alpha_i$ are chosen to cancel out all of the constant terms in
the equations of motion. This yields
\bea \alpha_{1} & = & -\frac{i\Omega}{\kappa}, \\ \alpha_2 & = &
-\frac{i\Omega}{2}\frac{\sqrt{2\zeta'}}{(\kappa/2)-i\delta_{2}'},
\eea
where $\delta_{2}'=\delta_{2}+2gkx_0$ is the detuning relative to
the new resonance frequency of the cavity when the sphere sits at
$x=x_0$ rather than $x=0$. Physically, $x=x_0$ corresponds to the
minimum of the total optical potential formed by the two driven
cavity modes. We define the ratio of the cavity mode intensities
to be $2\zeta\equiv|\alpha_{2}/\alpha_1|^2$, which is equivalent
to $\zeta=\zeta'\kappa^{2}/(\kappa^2+4\delta_{2}'^2)$. In terms of
$\zeta$, the shifted equilibrium position is given by
$kx_0=\zeta$. Clearly then the expansion in $\hat{x}$ of the
opto-mechanical coupling terms requires that $\zeta$ be small. For
simplicity, the prime symbol in $\delta_{2}'$ will be implicitly
understood, and we also take $\delta_1=0$ in the following
discussions. Following the shifts to the operators $\op{a}{i}$ and
$\hat{x}$ and then linearizing the equations of motion, one finds
\bea \frac{d}{dt}\op{a}{1} & = &
-4igk^{2}x_{0}\alpha_{1}\hat{x}-(\kappa/2)\op{a}{1}+\sqrt{\kappa}\op{a}{1,\footnotesize\textrm{in}},
\nonumber
\\ \frac{d}{dt}\op{a}{2} & = &
(i\delta_{2}-\kappa/2)\op{a}{2}+2ig\alpha_{2}k\hat{x}+\sqrt{\kappa}\op{a}{2,\footnotesize\textrm{in}},
\nonumber
\\ \frac{d}{dt}\hat{p} & = &
-4{\hbar}gk^{2}|\alpha_1|^2\hat{x}+2{\hbar}gk\left(\alpha_{2}\opdagger{a}{2}+\alpha_{2}^{\ast}\op{a}{2}-2kx_{0}(\alpha_{1}\opdagger{a}{1}+\alpha_{1}^{\ast}\op{a}{1})\right)-\gamma\hat{p}/2+\hat{F}_{p}(t),
\nonumber
\\ \frac{d}{dt}\hat{x} & = & \hat{p}/m.\label{eq:linearizedeqns} \eea
Note that cavity mode $1$ provides a linear restoring force
$d\hat{p}/dt{\sim}-4{\hbar}gk^{2}|\alpha_1|^{2}\hat{x}=-m\omega_{m}^{2}\hat{x}$,
and it is straightforward to show that this relation leads to the
expression for the harmonic oscillator frequency $\omega_m$ given
in Eq.~(1) of the main text. Furthermore, note that the sphere is
opto-mechanically coupled to mode $1$ with an amplitude
$4gk^{2}x_{0}\alpha_1{\propto}\zeta$, and to mode $2$ with an
amplitude $2gk\alpha_{2}{\propto}\sqrt{\zeta}$. Thus, to lowest
order in $\zeta$, modes $1$ and $2$ are purely responsible for
optical trapping and cooling, respectively. Treating mode $1$
simply as an external harmonic potential for the sphere, the
opto-mechanical system comprised of the CM motion of the sphere
and cavity mode $2$ is completely equivalent to the system
described in Ref.~\cite{wilson-rae07}. In particular, the optical
self-heating and self-cooling rates $R_{\pm}$ given in the main
text follow immediately. For convenience, we also re-define the
phases of the operators to make the opto-mechanical driving
amplitude
$\Omega_{m}=2{\hbar}gk\alpha_{2}=2{\hbar}gk\alpha_{1}\sqrt{2\zeta}$
real.

\section{Noise forces acting on trapped sphere}

In the main text, we have derived the motional heating rates of
the sphere due to background gas collisions and photon recoil
kicks, which under realistic conditions are the dominant heating
mechanisms. Here, we derive the heating rates for a number of
other less important processes.

\subsection{Photon shot noise}\label{subsec:shotnoise}

\begin{figure}[p]
\begin{center}
\includegraphics[width=10cm]{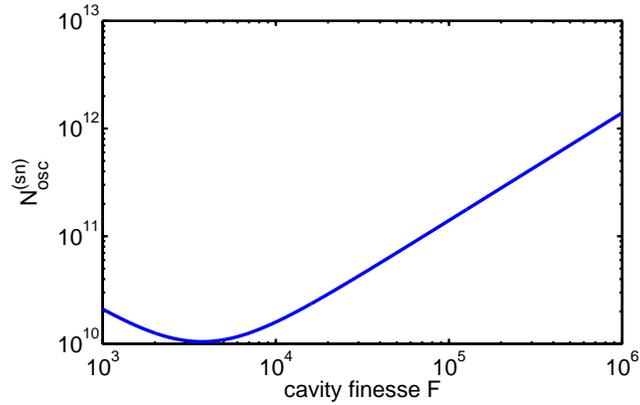}
\end{center}
\caption{The number of coherent oscillations
$N_{\footnotesize\textrm{osc}}$ allowed before a quantum jump due
to shot noise, as a function of cavity finesse. The system
parameters are given in the text.\label{fig:shotnoiseheating}}
\end{figure}

Photon shot noise inside the cavity leads to heating via
fluctuations in the mechanical oscillator frequency $\omega_m$. We
write the varying mechanical frequency in the form
\be
\omega_{m}^{2}(t)=\omega_{m,0}^2\left(1+\frac{\delta{N}(t)}{N_0}\right),
\ee
where $\omega_{m,0},N_0$ are the mean frequency and mean photon
number in the trapping mode of the cavity, and $\delta{N}$ is the
number fluctuation of this mode. Following the techniques of
Ref.~\cite{gehm98}, the shot noise leads to parametric
transitions~(where the phonon number $n{\rightarrow}n{\pm}2$ jumps
in pairs) at a rate $R$ proportional to the power spectral density
of the fluctuations at frequency $2\omega_{m,0}$,
\bea R_{n{\rightarrow}n+2} & = &
\frac{\pi\omega_{m,0}^2}{16}S(2\omega_{m,0})(n+2)(n+1), \\
R_{n{\rightarrow}n-2} & = &
\frac{\pi\omega_{m,0}^2}{16}S(2\omega_{m,0})n(n-1). \eea
Here the power spectral density is defined by
\be S(\omega)=\frac{2}{\pi
N_0^2}\int_{0}^{\infty}dt\;\cos\;{\omega}t\;\avg{\delta{N}(t)\delta{N}(0)},
\ee
which is evaluated to be
$S(\omega)=\frac{1}{{\pi}N_0}\frac{4\kappa}{\kappa^2+4\omega^2}$
for a cavity of linewidth $\kappa$ driven on resonance. Assuming
that the sphere initially is in the ground state, the number of
oscillations before a quantum jump due to shot noise is
\be
N_{\footnotesize\textrm{osc}}^{(\footnotesize\textrm{sn})}=\frac{\omega_{m,0}}{2{\pi}R_{0{\rightarrow}2}}=\frac{\epsilon+2}{\epsilon-1}\frac{V_{c}\rho}{3{\pi}c{\hbar}k^3}\frac{\omega_{m,0}}{\kappa}(\kappa^2+16\omega_{m,0}^2).
\ee
Here, $k=2\pi/\lambda$ is the wavevector of the trapping beam and
$V_c$ is the cavity mode volume. As an example, we consider a
cavity of length $L=1$~cm and waist
$w=25\;\mu$m~($V_c=(\pi/4)Lw^2$), a high-index dielectric
sphere~($\frac{\epsilon-1}{\epsilon+2}{\sim}1$), density
$\rho=2$~g/cm${}^3$, $\lambda=1\;\mu$m, and trapping frequency
$\omega_m/(2\pi)=0.5$~MHz.
$N_{\footnotesize\textrm{osc}}^{(\footnotesize\textrm{sn})}$ as a
function of cavity finesse $F$~($F=\pi{c}/2{\kappa}L$) is plotted
in Fig.~\ref{fig:shotnoiseheating}. It can be seen that the number
of allowed oscillations is at least of order
$N_{\footnotesize\textrm{osc}}^{(\footnotesize\textrm{sn})}{\sim}10^{10}$,
which is much larger than the limit due to photon recoil.
Physically, the low heating rates are attributable to the large
intra-cavity intensities used to achieve ${\sim}$MHz mechanical
oscillation frequencies, which suppresses the fractional noise
$\delta{N}/N_0{\propto}N_{0}^{-1/2}$.

\subsection{Blackbody radiation}

As in the case of scattering of laser light, the absorption and
emission of blackbody radiation by the sphere also lead to recoil
heating. The absorption rate of blackbody radiation of mode $\bfk$
is given in Eq.~(\ref{eq:Rabs})~(with each absorption event
providing a momentum kick ${\hbar}k_{x}$ along the trapping axis),
and again we assume that $\epsilon(\omega){\approx}\epsilon_{bb}$
is approximately flat across the blackbody radiation spectrum.
Summing over all modes, the characteristic jump rate due to
absorption of blackbody radiation is then given by~(cf. Eq.~(2) in
main text)
\be
\gamma_{\footnotesize\textrm{bb}}=\frac{2\pi^4}{63}\frac{(k_{B}T)^6}{c^5\hbar^5\rho\omega_m}\textrm{Im}\frac{\epsilon_{\footnotesize\textrm{bb}}-1}{\epsilon_{\footnotesize\textrm{bb}}+2}.
\ee
The jump rate between harmonic oscillator levels is
$R_{n{\rightarrow}n{\pm}1}=\gamma_{\footnotesize\textrm{bb}}(n+1/2{\pm}1/2)$.
An analogous expression holds for heating via the emission of
blackbody radiation, with the replacement
$T{\rightarrow}T_{\footnotesize\textrm{int}}$. Note that
$\gamma_{bb}$ is size-independent for small spheres, as both the
absorption rate and mass scale linearly with $V$. Taking as an
example a system with $\omega_{m}/(2\pi){\sim}1$~MHz,
$\rho=2$~g/cm${}^3$,
Im~$\frac{\epsilon_{\footnotesize\textrm{bb}}-1}{\epsilon_{\footnotesize\textrm{bb}}+2}=0.1$,
and $T{\sim}T_{\footnotesize\textrm{int}}{\sim}300$~K, we find
that the number of oscillations before a quantum jump~(due to
either absorption or emission) is
$N_{\footnotesize\textrm{osc}}^{(\footnotesize\textrm{bb})}{\sim}10^{11}$.

\subsection{Anisotropy of sphere}

The general problem of the rotational motion of an arbitrary
dielectric object inside an optical cavity is quite challenging to
solve. Generally, the polarizability
$\alpha_{\footnotesize\textrm{ind}}$ becomes a function of its
orientation, and changes in its orientation lead to changes in the
optical trapping potential and the intra-cavity intensity. Here we
consider a simplified version of the problem, where the rotational
motion is limited to one axis, and the anisotropy or deformation
of the sphere is of spheroid-type. As in the case of the sphere,
the latter assumption admits analytical solutions for the
polarizability tensor of the object~\cite{klimov02b}. In
particular, we assume that the dielectric is a prolate
nanospheroid whose size is much smaller than the optical
wavelength, with semi-major axis $a$ and semi-minor axis $b$, and
that the ratio $a/b{\approx}1$~(\textit{i.e.}, the deviation from
an ideal sphere is small).  Then the polarizability of the
spheroid is given by
\be
\alpha_{\footnotesize\textrm{ind}}{\approx}\alpha_{\footnotesize\textrm{ind},0}\left(1{\pm}\frac{9}{20}\frac{\epsilon-1}{\epsilon+2}\left[(a/b)^{4/3}-1\right]\right)\label{eq:spheroidalpha}
\ee
with
$\alpha_{\footnotesize\textrm{ind},0}{\approx}3\epsilon_{0}V\frac{\epsilon-1}{\epsilon+2}$.
Here the $\pm$ symbols denote when the major and minor axes are
aligned along the field polarization axis, respectively. From
Eqs.~(\ref{eq:cavityshift}) and~(\ref{eq:spheroidalpha}), it is
straightforward to find the shift in the cavity frequency taking
into account the rotational degree of freedom,
\be
\delta\omega=\delta\omega_0+\delta\omega_{\theta}\cos\,2\theta,
\ee
where $\delta\omega_0$ is the shift associated with the CM
position alone~(as given by Eq.~(\ref{eq:cmshift})), and
\be
\delta\omega_{\theta}=\frac{27}{80}\frac{V}{V_c}\left(\frac{\epsilon-1}{\epsilon+2}\right)^{2}\left[(a/b)^{4/3}-1\right]\omega\cos(2kx-2\phi).
\ee
Here we have defined $\theta$ as the angle of rotation of the
spheroid.

We are now interested in deriving the effect of the rotational
motion on the CM motion. In analogy with
Eq.~(\ref{eq:heisenbergeqs}), the coupled equations of motion
between the rotation and the trapping mode are
\bea \frac{da_1}{dt} & = &
-\left(i\delta\omega_{\theta}\cos\,2\theta+\frac{\kappa}{2}\right)a_1+\frac{i\Omega}{2},
\nonumber \\ \frac{dp_{\theta}}{dt} & = &
2\hbar\delta\omega_{\theta}|a_1|^{2}\sin\,2\theta-\gamma_{\theta}p_{\theta}+F_{\theta}(t),
\nonumber \\ \frac{d\theta}{dt} & = &
\frac{p_{\theta}}{I_{\theta}},\label{eq:nonlinearrotation} \eea
where $p_{\theta}$ is the angular momentum associated with
$\theta$, $I_{\theta}$ is the moment of inertia, and
$\gamma_{\theta},F_{\theta}$ are the damping coefficient and noise
force acting on the rotational motion. Since the rotational energy
is of order ${\sim}k_{B}T$, it suffices to consider the classical
equations given here.  The damping term is effected through the
background gas, as each collision between the spheroid and a gas
molecule partly exchanges angular momentum between the two
systems. The damping coefficient is found to be
$\gamma_{\theta}=5\sqrt{3/(2\pi)}\alpha_{\theta}P/(v_{\footnotesize\textrm{rms}}r\rho)$~\cite{volkov09},
where $r{\approx}a{\approx}b$. $\alpha_{\theta}$ is a
phenomenological accommodation coefficient describing the
efficiency of angular momentum transfer. The noise force has
correlations $\avg{F(t)F(t')}=2D\delta(t-t')$, where
$D=\gamma_{\theta}k_{B}T/I_{\theta}$. Note that $\gamma_{\theta}$
is a very small quantity under good vacuum conditions.

The full nonlinear coupled equations of
Eq.~(\ref{eq:nonlinearrotation}) are difficult to treat in a
general setting. However, given the typical smallness of the
parameters $\delta\omega_{\theta}/\kappa$ and
$\hbar\delta\omega_{\theta}|a_1|^2/(k_{B}T)$ for nearly spherical
particles, to lowest order we can ignore the optical coupling to
the rotational motion, and the dominant effect of the sphere
anisotropy is trap heating through fluctuations in the
polarizability $\alpha_{\footnotesize\textrm{ind}}$ rather than
intra-cavity intensity fluctuations. This leads to fluctuations in
the trap frequency given by
\be
\delta\omega_{m}(t)=\epsilon_{\theta}\omega_{m,0}\cos\,2\theta(t),
\ee
where
$\epsilon_{\theta}=\frac{9}{40}\frac{\epsilon-1}{\epsilon+2}((a/b)^{4/3}-1)$.
In analogy with the discussion in Sec.~\ref{subsec:shotnoise},
these fluctuations lead to parametric heating, with a jump rate
out of the ground state given by
\be
R_{0{\rightarrow}2}=\int_{0}^{\infty}dt\,\cos\,2\omega_{m,0}t\,\avg{\delta\omega_{m}(0)\delta\omega_{m}(t)}.
\ee
Denoting $\delta\theta(t)=\theta(t)-\theta(0)$, the above equation
can be re-written in the form
\be
R_{0{\rightarrow}2}=\frac{1}{2}\int_{0}^{\infty}dt\,\cos\,2\omega_{m,0}t\,(\epsilon_{\theta}\omega_{m,0}^2)^{2}\avg{\cos\,2\delta\theta(t)}.
\ee
Making a Gaussian approximation
$\avg{e^{2i\delta\theta(t)}}{\approx}\exp(-\avg{\delta\theta^{2}(t)}/2)$,
and taking the limit of small $\gamma_{\theta}$, one finally finds
\be
\frac{R_{0{\rightarrow}2}}{\omega_{m,0}}=\epsilon_{\theta}^{2}\frac{\sqrt{2\pi}\omega_{m,0}}{8\sqrt{\avg{\omega_r^2}}}\exp\left(-\frac{\omega_{m,0}^2}{2\avg{\omega_r^2}}\right).
\ee
Here $\omega_{r}=d\theta/dt$ is the angular velocity of the
spheroid~(typical values of $\sqrt{\avg{\omega_r^2}}$ are in the
MHz range for sub-wavelength particles). Note that the above
function is peaked at $\omega_{m,0}=\sqrt{\avg{\omega_r^2}}$,
\textit{i.e.}, the parametric heating is most pronounced when the
rotational frequency is comparable to the CM oscillation
frequency. At this maximum,
$R_{0{\rightarrow}2}/\omega_{m,0}{\sim}0.2\epsilon_{\theta}^2$.
Furthermore, for this worst-case scenario,
$R_{0{\rightarrow}2}/\omega_{m,0}$ can be suppressed to the
${\sim}10^{-5}$ level with an anisotropy of $a/b{\sim}1.03$.

\section{Analysis of entanglement transfer}\label{sec:EPR}

Here we provide a detailed analysis of entanglement transfer
between two modes of light and two spatially separate spheres,
leading to Eq.~(8) in the main text. The EPR correlations between
the two light modes given by Eq.~(7) in the main text,
\be
\avg{(X^{(A)}_{+,\footnotesize\textrm{in}}(\omega)+X^{(B)}_{+,\footnotesize\textrm{in}}(\omega))^2}/2=\avg{(X^{(A)}_{-,\footnotesize\textrm{in}}(\omega)-X^{(B)}_{-,\footnotesize\textrm{in}}(\omega))^2}/2=e^{-2R}<1,
\ee
are of the form created by a non-degenerate optical parametic
amplifier~(NOPA)~\cite{ou92}, which we describe below.

The Hamiltonian corresponding to a NOPA with cavity modes $A,B$ is
given by
\be
H=i\hbar(\beta/2)(\hat{c}^{(A)}\hat{c}^{(B)}-\hat{c}^{(A)\dagger}\hat{c}^{(B)\dagger}),
\ee
where $\hat{c}^{(j)}$ is the annihilation operator of mode $j$.
Taking an ideal, one-sided cavity~\cite{gardiner85}, and assuming
that the modes have identical linewidths $\kappa_c$, the
Heisenberg equations of motion for each mode read
\be
\frac{d}{dt}\hat{c}^{(j)}=-\frac{\kappa_c}{2}\hat{c}^{(j)}-\frac{\beta}{2}\hat{c}^{(j')\dagger}+\sqrt{\kappa_c}\hat{c}^{(j)}_{\footnotesize\textrm{in}}.\label{eq:NOPAevolution}
\ee
Here $\hat{c}^{(j)}_{\footnotesize\textrm{in}}$ is the cavity
input field for mode $j$, and $j'=A,B$ for $j=B,A$. The output
field is related to the intra-cavity and input fields by
$\hat{c}^{(j)}_{\footnotesize\textrm{out}}=\sqrt{\kappa_c}\hat{c}^{(j)}-\hat{c}^{(j)}_{\footnotesize\textrm{in}}$.
Writing
$\hat{c}^{(j)}(t)=(1/\sqrt{2\pi})\int\,d\omega\,e^{-i{\omega}t}\hat{c}^{(j)}(\omega)$,
Eq.~(\ref{eq:NOPAevolution}) can be exactly solved in the Fourier
domain for $\hat{c}^{(j)}(\omega)$. Specifically, defining
quadrature operators
$\hat{X}_{+}^{(j)}=\hat{c}^{(j)}+\hat{c}^{(j)\dagger}$ and
$\hat{X}_{-}^{(j)}=(\hat{c}^{(j)}-\hat{c}^{(j)\dagger})/i$~(with
analogous definitions for the quadrature operators of the input
and output fields), one can show that
\be
\hat{X}_{\pm,\footnotesize\textrm{out}}^{(A)}(\omega)\pm\hat{X}_{\pm,\footnotesize\textrm{out}}^{(B)}(\omega)=\frac{\kappa_c-\beta+2i\omega}{\kappa_c+\beta-2i\omega}\left(\hat{X}_{\pm,\footnotesize\textrm{in}}^{(A)}(\omega)\pm\hat{X}_{\pm,\footnotesize\textrm{in}}^{(B)}(\omega)\right).\;\;\;\;\;(\beta<\kappa_c)\ee
Over a bandwidth $\Delta\omega{\ll}\kappa_c$ that is much smaller
than the cavity linewidth, one can ignore the $\omega$ dependence
in the equation above, yielding
\be
\hat{X}_{\pm,\footnotesize\textrm{out}}^{(A)}(\omega)\pm\hat{X}_{\pm,\footnotesize\textrm{out}}^{(B)}(\omega)=e^{-R}\left(\hat{X}_{\pm,\footnotesize\textrm{in}}^{(A)}(\omega)\pm\hat{X}_{\pm,\footnotesize\textrm{in}}^{(B)}(\omega)\right),
\ee
where $e^{-R}=\frac{\kappa_c-\beta}{\kappa_c+\beta}$ for
$\beta<\kappa_c$. Physically, for non-zero $\beta$, the joint
variance of these quadratures in the output fields can display
reduced fluctuations relative to the input fields. It can also be
shown that the other combinations of the quadratures~(for
$\Delta\omega{\ll}\kappa_c$) satisfy
\be
\hat{X}_{\pm,\footnotesize\textrm{out}}^{(A)}(\omega)\mp\hat{X}_{\pm,\footnotesize\textrm{out}}^{(B)}(\omega)=e^{R}\left(\hat{X}_{\pm,\footnotesize\textrm{in}}^{(A)}(\omega)\mp\hat{X}_{\pm,\footnotesize\textrm{in}}^{(B)}(\omega)\right),
\ee
such that their joint variances become enhanced. For this
discussion, the input fields to the NOPA are assumed to be vacuum
states.

We now consider the quantum state transfer process for two spheres
trapped in spatially separate cavities, where the two output
fields generated by NOPA are fed as input fields into each of the
opto-mechanical systems. The equations of motion for the two
opto-mechanical systems~(denoted $A,B$) are given by Eq.~(6) in
the main text, with the replacement
$\hat{a}_{2,\footnotesize\textrm{in}}^{(j)}=\hat{c}_{\footnotesize\textrm{out}}^{(j)}$.
As in the main text, for simplicity we suppress the subscript
``2'' in the field operators denoting the trapping mode, since we
are only interested in this mode from this point on. To solve
these equations, we again work in the Fourier domain. Without the
fast-rotating terms $e^{2i\omega_{m}t}$, one could achieve ideal
state transfer between the mechanical motion and light, as
discussed in the main text. When the fast-rotating terms
$e^{2i\omega_{m}t}$ are included in the analysis, the frequency
components $\omega,\omega+2n\omega_{m}$~(integer $n$) of the
operators are coupled together in an infinite set of algebraic
equations. To make the problem tractable, we truncate this
infinite set by ignoring the components
$\hat{a}^{(j)}(\omega+2n\omega_m),\hat{b}^{(j)}(\omega+2n\omega_m)$
where $|n|{\geq}2$~(\textit{e.g.}, we assume
$\hat{a}^{(j)}(\omega{\pm}4\omega_m)=0$). This truncation
essentially amounts to the assumption that $\omega_{m}$ is large
compared to the other frequency scales in the problem. We then
solve the coupled set of equations for
$\hat{a}^{(j)}(\omega),\hat{b}^{(j)}(\omega)$ in terms of
$\hat{F}^{(j)}(\omega)$ and
$\hat{a}^{(j)}_{\footnotesize\textrm{in}}(\omega)$~(or
$\hat{c}^{(j)}_{\footnotesize\textrm{in}}(\omega)$), which allows
us to obtain any correlation functions for the cavity field or
mechanical motion in terms of those of the noise and input fields.
The noise forces $\hat{F}^{(j)}$ are assumed to be dominated by
photon recoil heating and are independent for the systems $A,B$,
such that their correlations take the form
$\avg{\hat{F}^{(j)}(\omega)}=0$ and
$\avg{\hat{F}^{(j)}(\omega)\hat{F}^{(j')}(\omega')}=\phi\omega_m\delta(\omega+\omega')\delta_{jj'}$,
where
$\phi=(4\pi^2/5)(V/\lambda^3)\frac{\epsilon-1}{\epsilon+2}$~(see
main text). We are specifically interested in the quantity
\be
\Delta_{\footnotesize\textrm{EPR}}\equiv\avg{(X^{(A)}_{\pm,m}(t){\mp}X^{(B)}_{\pm,m}(t))^2}/2
\ee
characterizing the joint variance in the motion of the two
spheres. The solution is generally quite complicated, but can be
expanded to lowest order in the small parameter
$\kappa/\omega_m$~(it is reasonably assumed that sideband
resolution can be achieved, so that $\kappa/\omega_m{\ll}1$).
After performing this procedure, and also ignoring any
fast-rotating terms~($e^{{\pm}2i\omega_{m}t}$) in the final
expression for $\Delta_{\footnotesize\textrm{EPR}}$, one arrives
at the solution given by Eq.~(8) in the main text.

\section{Analysis of squeezed light generation}

Here we derive the squeezing amplitude given in Eq.~(9) of the
main text. In the main text, it was argued that the trapping mode
of the cavity can be effectively considered as a mechanical
potential in the limit of small $\zeta$. We consider the situation
where the trapping beam intensity is varied to produce a
sinusoidal component in the mechanical spring constant at
frequency $2\omega_m$, with an effective Hamiltonian for the
motion given by
\bea H_{m} & = &
\frac{\hat{p}^2}{2m}+\frac{1}{2}m\omega_m^{2}\hat{x}^2(1+2\epsilon_{m}\sin\,2\omega_{m}t)
\\ & = & \hbar\omega_{m}\opdagger{b}{}\hat{b}-i\frac{\hbar\beta}{2}(\hat{b}^{2}e^{2i\omega_{m}t}-\opdagger{b}{}^{2}e^{-2i\omega_{m}t})+2\left\{\hbar\beta\opdagger{b}{}\hat{b}\sin\,2\omega_{m}t\right\}. \eea
In the last line, we have re-written
$\hat{x}=\sqrt{\frac{\hbar}{2m\omega_m}}(\hat{b}+\opdagger{b}{})$
and
$\hat{p}=i\sqrt{\frac{{\hbar}m\omega_m}{2}}(\opdagger{b}{}-\hat{b})$
in terms of the harmonic oscillator annihilation operator
$\hat{b}$ and also defined $\beta=\epsilon\omega_{m}/2$~(unrelated
to the $\beta$ term defined in the previous section for a NOPA).
The term in braces is a fast-varying contribution to the
Hamiltonian, in addition to the ``ideal'' squeezing Hamiltonian
comprising the remaining terms. The external Hamiltonian
$H_e$~(see Eq.~(6) in the main text) in this case is
\be
H_e=-i\frac{\hbar\beta}{2}(\hat{b}^{2}e^{2i\omega_{m}t}-\opdagger{b}{}^{2}e^{-2i\omega_{m}t})+2\hbar\beta\opdagger{b}{}\hat{b}\sin\,2\omega_{m}t,
\ee
while the Heisenberg equations of motion read
\bea \frac{d}{dt}\op{a}{2} & = &
-\frac{\kappa}{2}\op{a}{2}-i\Omega_{m}\left(\hat{b}+\opdagger{b}{2}e^{2i\omega_{m}t}\right)+\sqrt{\kappa}\op{a}{2,\footnotesize\textrm{in}},
\nonumber \\ \frac{d}{dt}\hat{b} & = &
-i\Omega_{m}\left(\op{a}{2}+\opdagger{a}{2}e^{2i\omega_{m}t}\right)+i\hat{F}(t)e^{i\omega_{m}t}+\beta\opdagger{b}{}-2i\beta\hat{b}\sin\,2\omega_{m}t.\label{eq:opto2}
\eea
We proceed to solve these equations in the Fourier domain using
the same techniques described in Sec.~\ref{sec:EPR}. Specifically,
we truncate terms containing frequency components
$\omega+2n\omega_{m}$~(integer $n$) at $|n|{\geq}2$ and solve for
$\hat{a}(\omega),\hat{b}(\omega)$ in terms of $\hat{F}(\omega)$
and $\hat{a}_{\footnotesize\textrm{in}}(\omega)$, from which any
correlation functions for the cavity field or mechanical motion
can be obtained. The input field is assumed to be in the vacuum
state. Similarly, the properties of the output field can be
obtained from these solutions by using the relation
$\op{a}{\footnotesize\textrm{out}}=\sqrt{\kappa}\op{a}{}-\op{a}{\footnotesize\textrm{in}}$.

We are specifically interested in the properties of the operator
$X_{+,\footnotesize\textrm{out}}(\omega=0)=\op{a}{\footnotesize\textrm{out}}(\omega=0)+\opdagger{a}{\footnotesize\textrm{out}}(\omega=0)$.
The general solutions of Eq.~(\ref{eq:opto2}) in the Fourier
domain are quite cumbersome, so we consider the simplified limit
where we set $\Gamma=\kappa$, and take the parametric driving
strength to be $\beta=\frac{\Gamma}{2}(1-\delta_t)$, where
$\delta_t{\ll}1$ is a small parameter that characterizes how far
one operates from threshold~($\beta{\rightarrow}\Gamma/2$).
Expanding to lowest order in $\kappa/\omega_m$ and $\delta_t$ and
ignoring fast-rotating terms, we find the following variance,
\be
{\Delta}X_{+,\footnotesize\textrm{out}}^2(\omega=0){\approx}\frac{5}{16}\frac{\kappa^2}{\omega_m^2}+\frac{3}{32}\frac{\kappa^2}{\omega_m^2}\delta_t+\frac{2\phi\omega_m}{\kappa}(1+\delta_t)+\frac{\delta_t^2}{4}.\label{eq:squeezing2}
\ee
In particular, at threshold~($\delta_t=0$), one recovers Eq.~(9)
of the main text. Maximum squeezing of the variance on threshold
is achieved when $\kappa=2(2\phi/5)^{1/3}\omega_m$, in which case
$({\Delta}X^{2}_{+,\footnotesize\textrm{out}})_{\footnotesize\textrm{min}}=(3/2)(5\phi^2/2)^{1/3}$.

Now we consider the effects of cavity loss on the maximum
achievable squeezing of output light. Starting from
Eq.~(\ref{eq:squeezing}) for the squeezing at threshold in an
ideal cavity~(with $\Gamma{\sim}\kappa$), we model cavity losses
via a beam splitter transformation with the ideal squeezed light
and vacuum as the two inputs. The output light exhibits reduced
squeezing due to mixing with the vacuum, given by
\be
({\Delta}X_{+,\footnotesize\textrm{out}}^2(\omega=0))_{\footnotesize\textrm{min}}=\left(1-\frac{\kappa'}{\kappa}\right)\left(\frac{2\phi\omega_m}{
\kappa}+\frac{5}{16}\frac{\kappa^2}{\omega_m^2}\right)+\frac{\kappa'}{\kappa},
\ee
where $\kappa',\kappa$ denote the scattering/absorption loss in
the cavity and the total cavity linewidth, respectively. In the
relevant regime where $\kappa'/\kappa{\ll}1$, we can approximate
$1-\kappa'/\kappa{\approx}1$ and the squeezing is optimized for
the choice $\kappa=2(2/5)^{1/3}(\phi+\kappa'/(2\omega_m))^{1/3}$,
for which
$({\Delta}X^{2}_{+,\footnotesize\textrm{out}})_{\footnotesize\textrm{min}}{\approx}
2.04(\phi+\kappa'/(2\omega_m))^{2/3}$. We now must choose a set of
realistic cavity parameters where this optimized squeezing can be
realized, and where $\Gamma{\sim}\kappa$ is consistent with
$\zeta$ being small. As an example, taking a cavity length and
waist of $L{\sim}2$~cm and $w{\sim}10\;\mu$m, $\kappa'$
corresponding to $1$~ppm losses per round trip, and sphere
parameters $r{\sim}35$~nm and $\omega_{m}/(2\pi){\sim}0.65$~MHz,
we find that $\Gamma{\sim}\kappa$ corresponds to a value
$\zeta{\sim}1/4$, which yields squeezing of $~{\sim}15$~dB in the
output light.

\begin{figure}[p]
\begin{center}
\includegraphics[width=10cm]{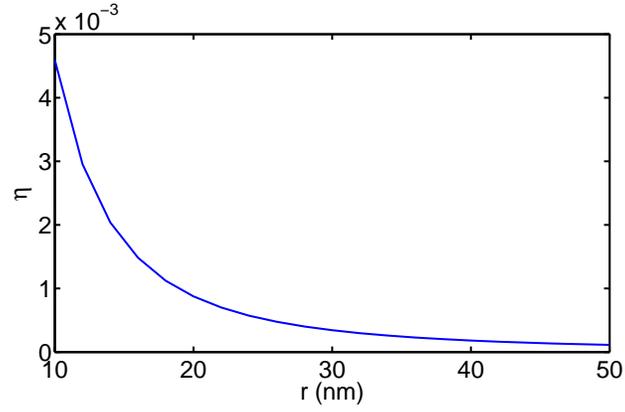}
\end{center}
\caption{The Lamb-Dicke parameter $\eta=k\Delta{x}$ corresponding
to the squeezed motional state of the sphere, as a function of
sphere size. The squeezing parameters are chosen such that the
squeezing in the output light is increased by 1~dB over
$({\Delta}X^{2}_{+,\footnotesize\textrm{out}})_{\footnotesize\textrm{min}}$.
The physical parameters of the system are taken to be
$\lambda=1\;\mu$m, $\rho=2$~g/cm${}^3$,
$\frac{\epsilon-1}{\epsilon+2}{\sim}1$, and
$\omega_{m}/(2\pi)=1$~MHz.\label{fig:lambdicke}}
\end{figure}

Thus far, we have neglected to consider corrections due to a
possibly large position uncertainty $\Delta{x}$ for the CM motion
of the sphere. Specifically, as one approaches threshold, one
quadrature of motion becomes infinitely unsqueezed, producing a
large ${\Delta}x$. At the same time, faithful quantum state
transfer requires a linear opto-mechanical coupling, where
$\mathcal{O}(x^2)$ shifts in the cavity cooling mode frequency can
be ignored. Specifically, the Lamb-Dicke parameter
$\eta{\equiv}k{\Delta}x{\ll}1$ for the trapped sphere must remain
small. To quantify this effect, we consider the situation where we
operate away from threshold by an amount that decreases the
squeezing by just $1$~dB relative to
$({\Delta}X^{2}_{+,\footnotesize\textrm{out}})_{\footnotesize\textrm{min}}$.
The value of $\delta_t$ corresponding to this 1~dB increase can be
obtained by solving Eq.~(\ref{eq:squeezing2}), and plugged into
the solutions of Eq.~(\ref{eq:opto2}) to numerically find
$\Delta{x}$. For concreteness, here we associate $\Delta{x}$ with
the position uncertainty in the unsqueezed quadrature of motion.
The corresponding Lamb-Dicke parameter as a function of sphere
size is then plotted in Fig.~\ref{fig:lambdicke} for the choice
$\omega_{m}/(2\pi)=1$~MHz, and it is seen that $\eta<10^{-2}$ over
the entire parameter regime.

\bibliographystyle{naturemag}
\bibliography{../bibsalpha}

\begin{thebibliography}{10}
\expandafter\ifx\csname url\endcsname\relax
  \def\url#1{\texttt{#1}}\fi
\expandafter\ifx\csname urlprefix\endcsname\relax\def\urlprefix{URL }\fi
\providecommand{\bibinfo}[2]{#2}
\providecommand{\eprint}[2][]{\url{#2}}

\bibitem{cleland09}
\bibinfo{author}{Cleland, A.}
\newblock \bibinfo{title}{{Optomechanics: Photons refrigerating phonons}}.
\newblock \emph{\bibinfo{journal}{Nature Phys.}} \textbf{\bibinfo{volume}{5}},
  \bibinfo{pages}{458--460} (\bibinfo{year}{2009}).

\bibitem{mamin01}
\bibinfo{author}{Mamin, H.~J.} \& \bibinfo{author}{Rugar, D.}
\newblock \bibinfo{title}{Sub-attonewton force detection at millikelvin
  temperatures}.
\newblock \emph{\bibinfo{journal}{Appl. Phys. Lett.}}
  \textbf{\bibinfo{volume}{79}}, \bibinfo{pages}{3358--3360}
  (\bibinfo{year}{2001}).

\bibitem{cleland04}
\bibinfo{author}{Cleland, A.~N.} \& \bibinfo{author}{Geller, M.~R.}
\newblock \bibinfo{title}{{Superconducting Qubit Storage and Entanglement with
  Nanomechanical Resonators}}.
\newblock \emph{\bibinfo{journal}{Phys. Rev. Lett.}}
  \textbf{\bibinfo{volume}{93}}, \bibinfo{pages}{070501}
  (\bibinfo{year}{2004}).

\bibitem{hao03}
\bibinfo{author}{Hao, Z.}, \bibinfo{author}{Erbil, A.} \&
  \bibinfo{author}{Ayazi, F.}
\newblock \bibinfo{title}{An analytical model for support loss in micromachined
  beam resonators with in-plane flexural vibrations}.
\newblock \emph{\bibinfo{journal}{Sens. Actuators A}}
  \textbf{\bibinfo{volume}{109}}, \bibinfo{pages}{156 -- 164}
  (\bibinfo{year}{2003}).

\bibitem{lifshitz00}
\bibinfo{author}{Lifshitz, R.} \& \bibinfo{author}{Roukes, M.~L.}
\newblock \bibinfo{title}{Thermoelastic damping in micro- and nanomechanical
  systems}.
\newblock \emph{\bibinfo{journal}{Phys. Rev. B}} \textbf{\bibinfo{volume}{61}},
  \bibinfo{pages}{5600--5609} (\bibinfo{year}{2000}).

\bibitem{ashkin07}
\bibinfo{author}{Ashkin, A.}
\newblock \emph{\bibinfo{title}{{Optical Trapping and Manipulation of Neutral
  Particles Using Lasers: a Reprint Volume with Commentaires}}}
  (\bibinfo{publisher}{World Scientific}, \bibinfo{year}{2007}).

\bibitem{ashkin76}
\bibinfo{author}{Ashkin, A.} \& \bibinfo{author}{Dziedzic, J.~M.}
\newblock \bibinfo{title}{Optical levitation in high vacuum}.
\newblock \emph{\bibinfo{journal}{Appl. Phys. Lett.}}
  \textbf{\bibinfo{volume}{28}}, \bibinfo{pages}{333--335}
  (\bibinfo{year}{1976}).

\bibitem{libbrecht04}
\bibinfo{author}{Libbrecht, K.~G.} \& \bibinfo{author}{Black, E.~D.}
\newblock \bibinfo{title}{{Toward quantum-limited position measurements using
  optically levitated microspheres}}.
\newblock \emph{\bibinfo{journal}{Phys. Lett. A}}
  \textbf{\bibinfo{volume}{321}}, \bibinfo{pages}{99--102}
  (\bibinfo{year}{2004}).

\bibitem{braginsky02}
\bibinfo{author}{{Braginsky}, V.~B.} \& \bibinfo{author}{{Vyatchanin}, S.~P.}
\newblock \bibinfo{title}{{Low quantum noise tranquilizer for Fabry-Perot
  interferometer}}.
\newblock \emph{\bibinfo{journal}{Phys. Lett. A}}
  \textbf{\bibinfo{volume}{293}}, \bibinfo{pages}{228--234}
  (\bibinfo{year}{2002}).

\bibitem{wilson-rae07}
\bibinfo{author}{Wilson-Rae, I.}, \bibinfo{author}{Nooshi, N.},
  \bibinfo{author}{Zwerger, W.} \& \bibinfo{author}{Kippenberg, T.~J.}
\newblock \bibinfo{title}{{Theory of Ground State Cooling of a Mechanical
  Oscillator Using Dynamical Backaction}}.
\newblock \emph{\bibinfo{journal}{Phys. Rev. Lett.}}
  \textbf{\bibinfo{volume}{99}}, \bibinfo{pages}{093901}
  (\bibinfo{year}{2007}).

\bibitem{marquardt07}
\bibinfo{author}{Marquardt, F.}, \bibinfo{author}{Chen, J.~P.},
  \bibinfo{author}{Clerk, A.~A.} \& \bibinfo{author}{Girvin, S.~M.}
\newblock \bibinfo{title}{{Quantum Theory of Cavity-Assisted Sideband Cooling
  of Mechanical Motion}}.
\newblock \emph{\bibinfo{journal}{Phys. Rev. Lett.}}
  \textbf{\bibinfo{volume}{99}}, \bibinfo{pages}{093902}
  (\bibinfo{year}{2007}).

\bibitem{zeng94}
\bibinfo{author}{Zeng, H.} \& \bibinfo{author}{Lin, F.}
\newblock \bibinfo{title}{Quantum conversion between the cavity fields and the
  center-of-mass motion of ions in a quantized trap}.
\newblock \emph{\bibinfo{journal}{Phys. Rev. A}} \textbf{\bibinfo{volume}{50}},
  \bibinfo{pages}{R3589--R3592} (\bibinfo{year}{1994}).

\bibitem{parkins99}
\bibinfo{author}{{Parkins}, A.~S.} \& \bibinfo{author}{{Kimble}, H.~J.}
\newblock \bibinfo{title}{{Quantum state transfer between motion and light}}.
\newblock \emph{\bibinfo{journal}{J. Opt. B}} \textbf{\bibinfo{volume}{1}},
  \bibinfo{pages}{496--504} (\bibinfo{year}{1999}).

\bibitem{zhang03}
\bibinfo{author}{Zhang, J.}, \bibinfo{author}{Peng, K.} \&
  \bibinfo{author}{Braunstein, S.~L.}
\newblock \bibinfo{title}{Quantum-state transfer from light to macroscopic
  oscillators}.
\newblock \emph{\bibinfo{journal}{Phys. Rev. A}} \textbf{\bibinfo{volume}{68}},
  \bibinfo{pages}{013808} (\bibinfo{year}{2003}).

\bibitem{jahne09}
\bibinfo{author}{J\"{a}hne, K.} \emph{et~al.}
\newblock \bibinfo{title}{Cavity-assisted squeezing of a mechanical
  oscillator}.
\newblock \emph{\bibinfo{journal}{Phys. Rev. A}} \textbf{\bibinfo{volume}{79}},
  \bibinfo{pages}{063819} (\bibinfo{year}{2009}).

\bibitem{wu87}
\bibinfo{author}{Wu, L.-A.}, \bibinfo{author}{Xiao, M.} \&
  \bibinfo{author}{Kimble, H.~J.}
\newblock \bibinfo{title}{Squeezed states of light from an optical parametric
  oscillator}.
\newblock \emph{\bibinfo{journal}{J. Opt. Soc. Am. B}}
  \textbf{\bibinfo{volume}{4}}, \bibinfo{pages}{1465--1475}
  (\bibinfo{year}{1987}).

\bibitem{yonezawa07}
\bibinfo{author}{Yonezawa, H.}, \bibinfo{author}{Braunstein, S.~L.} \&
  \bibinfo{author}{Furusawa, A.}
\newblock \bibinfo{title}{{Experimental Demonstration of Quantum Teleportation
  of Broadband Squeezing}}.
\newblock \emph{\bibinfo{journal}{Phys. Rev. Lett.}}
  \textbf{\bibinfo{volume}{99}}, \bibinfo{pages}{110503}
  (\bibinfo{year}{2007}).

\bibitem{einstein35}
\bibinfo{author}{Einstein, A.}, \bibinfo{author}{Podolsky, B.} \&
  \bibinfo{author}{Rosen, N.}
\newblock \bibinfo{title}{{Can Quantum-Mechanical Description of Physical
  Reality Be Considered Complete?}}
\newblock \emph{\bibinfo{journal}{Phys. Rev.}} \textbf{\bibinfo{volume}{47}},
  \bibinfo{pages}{777--780} (\bibinfo{year}{1935}).

\bibitem{mckeever03}
\bibinfo{author}{McKeever, J.} \emph{et~al.}
\newblock \bibinfo{title}{{State-Insensitive Cooling and Trapping of Single
  Atoms in an Optical Cavity}}.
\newblock \emph{\bibinfo{journal}{Phys. Rev. Lett.}}
  \textbf{\bibinfo{volume}{90}}, \bibinfo{pages}{133602}
  (\bibinfo{year}{2003}).

\bibitem{maunz04}
\bibinfo{author}{{Maunz}, P.} \emph{et~al.}
\newblock \bibinfo{title}{{Cavity cooling of a single atom}}.
\newblock \emph{\bibinfo{journal}{Nature}} \textbf{\bibinfo{volume}{428}},
  \bibinfo{pages}{50--52} (\bibinfo{year}{2004}).

\bibitem{leibfried03}
\bibinfo{author}{Leibfried, D.}, \bibinfo{author}{Blatt, R.},
  \bibinfo{author}{Monroe, C.} \& \bibinfo{author}{Wineland, D.}
\newblock \bibinfo{title}{Quantum dynamics of single trapped ions}.
\newblock \emph{\bibinfo{journal}{Rev. Mod. Phys.}}
  \textbf{\bibinfo{volume}{75}}, \bibinfo{pages}{281--324}
  (\bibinfo{year}{2003}).

\bibitem{jost09}
\bibinfo{author}{Jost, J.~D.} \emph{et~al.}
\newblock \bibinfo{title}{{Entangled mechanical oscillators}}.
\newblock \emph{\bibinfo{journal}{Nature}} \textbf{\bibinfo{volume}{459}},
  \bibinfo{pages}{683--685} (\bibinfo{year}{2009}).

\bibitem{murch08}
\bibinfo{author}{Murch, K.~W.}, \bibinfo{author}{Moore, K.~L.},
  \bibinfo{author}{Gupta, S.} \& \bibinfo{author}{Stamper-Kurn, D.~M.}
\newblock \bibinfo{title}{{Observation of quantum-measurement backaction with
  an ultracold atomic gas}}.
\newblock \emph{\bibinfo{journal}{Nature Phys.}} \textbf{\bibinfo{volume}{4}},
  \bibinfo{pages}{561--564} (\bibinfo{year}{2008}).

\bibitem{brennecke08}
\bibinfo{author}{Brennecke, F.}, \bibinfo{author}{Ritter, S.},
  \bibinfo{author}{Donner, T.} \& \bibinfo{author}{Esslinger, T.}
\newblock \bibinfo{title}{{Cavity Optomechanics with a Bose-Einstein
  Condensate}}.
\newblock \emph{\bibinfo{journal}{Science}} \textbf{\bibinfo{volume}{322}},
  \bibinfo{pages}{235--238} (\bibinfo{year}{2008}).

\bibitem{zener32}
\bibinfo{author}{Zener, C.}
\newblock \bibinfo{title}{{Non-adiabatic crossing of energy levels}}.
\newblock \emph{\bibinfo{journal}{Proc. R. Soc. London A}}
  \bibinfo{pages}{696--702} (\bibinfo{year}{1932}).

\bibitem{cirac97}
\bibinfo{author}{Cirac, J.~I.}, \bibinfo{author}{Zoller, P.},
  \bibinfo{author}{Kimble, H.~J.} \& \bibinfo{author}{Mabuchi, H.}
\newblock \bibinfo{title}{{Quantum State Transfer and Entanglement Distribution
  among Distant Nodes in a Quantum Network}}.
\newblock \emph{\bibinfo{journal}{Phys. Rev. Lett.}}
  \textbf{\bibinfo{volume}{78}}, \bibinfo{pages}{3221--3224}
  (\bibinfo{year}{1997}).

\bibitem{beveratos01}
\bibinfo{author}{Beveratos, A.}, \bibinfo{author}{Brouri, R.},
  \bibinfo{author}{Gacoin, T.}, \bibinfo{author}{Poizat, J.-P.} \&
  \bibinfo{author}{Grangier, P.}
\newblock \bibinfo{title}{Nonclassical radiation from diamond nanocrystals}.
\newblock \emph{\bibinfo{journal}{Phys. Rev. A}} \textbf{\bibinfo{volume}{64}},
  \bibinfo{pages}{061802} (\bibinfo{year}{2001}).

\bibitem{bahns96}
\bibinfo{author}{Bahns, J.~T.}, \bibinfo{author}{Stwalley, W.~C.} \&
  \bibinfo{author}{Gould, P.~L.}
\newblock \bibinfo{title}{{Laser cooling of molecules: A sequential scheme for
  rotation, translation, and vibration}}.
\newblock \emph{\bibinfo{journal}{J. Chem. Phys.}}
  \textbf{\bibinfo{volume}{104}}, \bibinfo{pages}{9689--9697}
  (\bibinfo{year}{1996}).

\bibitem{epstein24}
\bibinfo{author}{Epstein, P.~S.}
\newblock \bibinfo{title}{{On the Resistance Experienced by Spheres in their
  Motion through Gases}}.
\newblock \emph{\bibinfo{journal}{Phys. Rev.}} \textbf{\bibinfo{volume}{23}},
  \bibinfo{pages}{710--733} (\bibinfo{year}{1924}).

\bibitem{wineland79}
\bibinfo{author}{Wineland, D.~J.} \& \bibinfo{author}{Itano, W.~M.}
\newblock \bibinfo{title}{Laser cooling of atoms}.
\newblock \emph{\bibinfo{journal}{Phys. Rev. A}} \textbf{\bibinfo{volume}{20}},
  \bibinfo{pages}{1521--1540} (\bibinfo{year}{1979}).

\bibitem{cirac92}
\bibinfo{author}{Cirac, J.~I.}, \bibinfo{author}{Blatt, R.},
  \bibinfo{author}{Zoller, P.} \& \bibinfo{author}{Phillips, W.~D.}
\newblock \bibinfo{title}{Laser cooling of trapped ions in a standing wave}.
\newblock \emph{\bibinfo{journal}{Phys. Rev. A}} \textbf{\bibinfo{volume}{46}},
  \bibinfo{pages}{2668--2681} (\bibinfo{year}{1992}).

\bibitem{kokorowski01}
\bibinfo{author}{Kokorowski, D.~A.}, \bibinfo{author}{Cronin, A.~D.},
  \bibinfo{author}{Roberts, T.~D.} \& \bibinfo{author}{Pritchard, D.~E.}
\newblock \bibinfo{title}{{From single-to multiple-photon decoherence in an
  atom interferometer}}.
\newblock \emph{\bibinfo{journal}{Phys. Rev. Lett.}}
  \textbf{\bibinfo{volume}{86}}, \bibinfo{pages}{2191--2195}
  (\bibinfo{year}{2001}).

\bibitem{gardiner85}
\bibinfo{author}{Gardiner, C.~W.} \& \bibinfo{author}{Collett, M.~J.}
\newblock \bibinfo{title}{{Input and output in damped quantum systems: Quantum
  stochastic differential equations and the master equation}}.
\newblock \emph{\bibinfo{journal}{Phys. Rev. A}} \textbf{\bibinfo{volume}{31}},
  \bibinfo{pages}{3761--3774} (\bibinfo{year}{1985}).

\bibitem{duan00}
\bibinfo{author}{Duan, L.-M.}, \bibinfo{author}{Giedke, G.},
  \bibinfo{author}{Cirac, J.~I.} \& \bibinfo{author}{Zoller, P.}
\newblock \bibinfo{title}{{Inseparability Criterion for Continuous Variable
  Systems}}.
\newblock \emph{\bibinfo{journal}{Phys. Rev. Lett.}}
  \textbf{\bibinfo{volume}{84}}, \bibinfo{pages}{2722--2725}
  (\bibinfo{year}{2000}).

\bibitem{julsgaard01}
\bibinfo{author}{Julsgaard, B.}, \bibinfo{author}{Kozhekin, A.} \&
  \bibinfo{author}{Polzik, E.~S.}
\newblock \bibinfo{title}{{Experimental long-lived entanglement of two
  macroscopic objects}}.
\newblock \emph{\bibinfo{journal}{Nature}} \textbf{\bibinfo{volume}{413}},
  \bibinfo{pages}{400--403} (\bibinfo{year}{2001}).

\bibitem{rugar91}
\bibinfo{author}{Rugar, D.} \& \bibinfo{author}{Gr\"utter, P.}
\newblock \bibinfo{title}{Mechanical parametric amplification and
  thermomechanical noise squeezing}.
\newblock \emph{\bibinfo{journal}{Phys. Rev. Lett.}}
  \textbf{\bibinfo{volume}{67}}, \bibinfo{pages}{699--702}
  (\bibinfo{year}{1991}).

\bibitem{hood01}
\bibinfo{author}{Hood, C.~J.}, \bibinfo{author}{Kimble, H.~J.} \&
  \bibinfo{author}{Ye, J.}
\newblock \bibinfo{title}{{Characterization of high-finesse mirrors: Loss,
  phase shifts, and mode structure in an optical cavity}}.
\newblock \emph{\bibinfo{journal}{Phys. Rev. A}} \textbf{\bibinfo{volume}{64}},
  \bibinfo{pages}{033804} (\bibinfo{year}{2001}).

\bibitem{venkatathri08}
\bibinfo{author}{Venkatathri, N.} \& \bibinfo{author}{Yoo, J.~W.}
\newblock \bibinfo{title}{{Synthesis and Characterizationof Silica Nanosphere
  from Octadecyltrimethoxy Silane}}.
\newblock \emph{\bibinfo{journal}{Bull. Korean Chem. Soc.}}
  \textbf{\bibinfo{volume}{29}}, \bibinfo{pages}{29} (\bibinfo{year}{2008}).

\bibitem{shu06}
\bibinfo{author}{Shu, J.} \emph{et~al.}
\newblock \bibinfo{title}{Elastic light scattering from nanoparticles by
  monochromatic vacuum-ultraviolet radiation}.
\newblock \emph{\bibinfo{journal}{J. Chem. Phys.}}
  \textbf{\bibinfo{volume}{124}}, \bibinfo{pages}{034707}
  (\bibinfo{year}{2006}).

\bibitem{hackermuller04}
\bibinfo{author}{Hackerm\"{u}ller, L.}, \bibinfo{author}{Hornberger, K.},
  \bibinfo{author}{Brezger, B.}, \bibinfo{author}{Zeilinger, A.} \&
  \bibinfo{author}{Arndt, M.}
\newblock \bibinfo{title}{{Decoherence of matter waves by thermal emission of
  radiation}}.
\newblock \emph{\bibinfo{journal}{Nature}} \textbf{\bibinfo{volume}{427}},
  \bibinfo{pages}{711--714} (\bibinfo{year}{2004}).

\bibitem{vahala03}
\bibinfo{author}{{Vahala, K. J.}}
\newblock \bibinfo{title}{{Optical microcavities}}.
\newblock \emph{\bibinfo{journal}{Nature}} \textbf{\bibinfo{volume}{424}},
  \bibinfo{pages}{839--846} (\bibinfo{year}{2003}).

\bibitem{vuckovic03}
\bibinfo{author}{Vu{\v{c}}kovi{\'c}, J.} \& \bibinfo{author}{Yamamoto, Y.}
\newblock \bibinfo{title}{{Photonic crystal microcavities for cavity quantum
  electrodynamics with a single quantum dot}}.
\newblock \emph{\bibinfo{journal}{Appl. Phys. Lett.}}
  \textbf{\bibinfo{volume}{82}}, \bibinfo{pages}{2374--2376}
  (\bibinfo{year}{2003}).

\bibitem{wineland07}
\bibinfo{author}{Wineland, D.~J.} \emph{et~al.}
\newblock \bibinfo{title}{{Trapped atomic ions and quantum information
  processing}}.
\newblock In \bibinfo{editor}{Roos, C.}, \bibinfo{editor}{Haffner, H.} \&
  \bibinfo{editor}{Blatt, R.} (eds.) \emph{\bibinfo{booktitle}{{Proceedings of
  the International Conference on Atomic Physics (ICAP 2006)}}},
  \bibinfo{pages}{103--110} (\bibinfo{publisher}{American Institute of
  Physics}, \bibinfo{year}{2006}).

\bibitem{rabl09}
\bibinfo{author}{Rabl, P.} \emph{et~al.}
\newblock \bibinfo{title}{{A quantum spin transducer based on nano
  electro-mechancial resonator arrays}}.
\newblock \emph{\bibinfo{journal}{ArXiv e-prints}}  (\bibinfo{year}{2009}).
\newblock \eprint{0908.0316}.

\bibitem{stratton41}
\bibinfo{author}{Stratton, J.~A.}
\newblock \emph{\bibinfo{title}{{Electromagnetic theory, 1st ed.}}}
  (\bibinfo{publisher}{McGraw-Hill}, \bibinfo{address}{New York},
  \bibinfo{year}{1941}).

\bibitem{liu06}
\bibinfo{author}{{Liu}, F.}, \bibinfo{author}{{Daun}, K.~J.},
  \bibinfo{author}{{Snelling}, D.~R.} \& \bibinfo{author}{{Smallwood}, G.~J.}
\newblock \bibinfo{title}{{Heat conduction from a spherical nano-particle:
  status of modeling heat conduction in laser-induced incandescence}}.
\newblock \emph{\bibinfo{journal}{Appl. Phys. B}}
  \textbf{\bibinfo{volume}{83}}, \bibinfo{pages}{355--382}
  (\bibinfo{year}{2006}).

\bibitem{lang83}
\bibinfo{author}{Lang, M.~L.} \& \bibinfo{author}{Wolfe, W.~L.}
\newblock \bibinfo{title}{{Optical constants of fused silica and sapphire from
  0.3 to 25~$\mu$m}}.
\newblock \emph{\bibinfo{journal}{Appl. Opt.}} \textbf{\bibinfo{volume}{22}},
  \bibinfo{pages}{1267--1268} (\bibinfo{year}{1983}).

\bibitem{rodriguez07}
\bibinfo{author}{Rodriguez, A.}, \bibinfo{author}{Soljacic, M.},
  \bibinfo{author}{Joannopoulos, J.~D.} \& \bibinfo{author}{Johnson, S.~G.}
\newblock \bibinfo{title}{{$\chi^{(2)}$ and $\chi^{(3)}$ harmonic generation at
  a critical power in inhomogeneous doubly resonant cavities}}.
\newblock \emph{\bibinfo{journal}{Opt. Express}} \textbf{\bibinfo{volume}{15}},
  \bibinfo{pages}{7303--7318} (\bibinfo{year}{2007}).

\bibitem{gehm98}
\bibinfo{author}{Gehm, M.~E.}, \bibinfo{author}{O\char39{}Hara, K.~M.},
  \bibinfo{author}{Savard, T.~A.} \& \bibinfo{author}{Thomas, J.~E.}
\newblock \bibinfo{title}{Dynamics of noise-induced heating in atom traps}.
\newblock \emph{\bibinfo{journal}{Phys. Rev. A}} \textbf{\bibinfo{volume}{58}},
  \bibinfo{pages}{3914--3921} (\bibinfo{year}{1998}).

\bibitem{klimov02b}
\bibinfo{author}{Klimov, V.~V.}, \bibinfo{author}{Ducloy, M.} \&
  \bibinfo{author}{Letokhov, V.~S.}
\newblock \bibinfo{title}{Spontaneous emission of an atom placed near a prolate
  nanospheroid}.
\newblock \emph{\bibinfo{journal}{Eur. Phys. J. D}}
  \textbf{\bibinfo{volume}{20}}, \bibinfo{pages}{133--148}
  (\bibinfo{year}{2002}).

\bibitem{volkov09}
\bibinfo{author}{Volkov, A.~N.}
\newblock \bibinfo{title}{{Aerodynamic coefficients of a spinning sphere in a
  rarefied-gas flow}}.
\newblock \emph{\bibinfo{journal}{Fluid Dynamics}}
  \textbf{\bibinfo{volume}{44}}, \bibinfo{pages}{141--157}
  (\bibinfo{year}{2009}).

\bibitem{ou92}
\bibinfo{author}{Ou, Z.~Y.}, \bibinfo{author}{Pereira, S.~F.},
  \bibinfo{author}{Kimble, H.~J.} \& \bibinfo{author}{Peng, K.~C.}
\newblock \bibinfo{title}{{Realization of the Einstein-Podolsky-Rosen paradox
  for continuous variables}}.
\newblock \emph{\bibinfo{journal}{Phys. Rev. Lett.}}
  \textbf{\bibinfo{volume}{68}}, \bibinfo{pages}{3663--3666}
  (\bibinfo{year}{1992}).

\end{thebibliography}

\end{document}